\documentclass[reprint,amsmath,amssymb,aps,prc,nofootinbib]{revtex4-1}
\usepackage{bm}
\usepackage{graphicx}
\usepackage{dcolumn}
\usepackage{color}
\newcommand{\be}{\begin{equation}}
\newcommand{\ee}{\end{equation}}
\newcommand{\bea}{\begin{eqnarray}}
\newcommand{\eea}{\end{eqnarray}}

\newcommand{\bfl}{\mbox{\boldmath $l$}}

\newcommand{\bfbq}{\mbox{\boldmath $Q$}}

\newcommand{\bftau}{\mbox{\boldmath $\tau$}}
\newcommand{\bftaup}{\mbox{\boldmath $\tau'$}}

\newcommand{\mbss}[1]{_{\mbox{\scriptsize #1}}}

\newcommand{\mbsu}[1]{\mbox{\scriptsize #1}}

\newcommand{\vphu}{\vphantom{*}}
\newcommand{\vphd}{\vphantom{1}}
\newcommand{\hfm}{\hphantom{-}}
\newcommand{\bfsigma}{\mbox{\boldmath $\sigma$}}
\newcommand{\bfsigmap}{\mbox{\boldmath $\sigma'$}}

\newcommand{\ve}{\varepsilon}

\newcommand{\ffrac}[2]{{\textstyle\frac{#1}{#2}}}

\begin{document}

\title{
$M1$ resonance in $^{208}$Pb
within the self-consistent phonon-coupling model
}

\author{V. Tselyaev}
\email{tselyaev@mail.ru}
\author{N. Lyutorovich}
\affiliation{St. Petersburg State University, St. Petersburg, 199034, Russia}
\author{J. Speth}
\affiliation{Institut f\"ur Kernphysik, Forschungszentrum J\"ulich, D-52425 J\"ulich, Germany}
\author{P.-G. Reinhard}
\affiliation{Institut f\"ur Theoretische Physik II, Universit\"at Erlangen-N\"urnberg,
D-91058 Erlangen, Germany}
\date{\today}

\begin{abstract}
The main goal of the paper is to investigate theoretically  the
experimentally observed fragmentation of the isovector $M1$ resonance in $^{208}$Pb
within a self-consistent model based on an energy-density functional (EDF)
of the Skyrme type. This fragmentation (spread of the $M1$ strength) is not
reproduced in a conventional one-particle--one-hole ($1p1h$) random-phase approximation
(RPA) and thus has to be investigated in the framework of more complicated models.
However, previously applied models of this type were not self-consistent.
In the present work, we use a recently developed renormalized version of the self-consistent
time blocking approximation (RenTBA) in which the $1p1h\otimes$phonon configurations
are included on top of the RPA $1p1h$ configurations.
We have determined several sets of the parameters of the modified Skyrme EDF
fitted within the RenTBA and RPA and have found the necessary condition of
producing the fragmentation of the $M1$ resonance in $^{208}$Pb in our model.
We present also the results of the RenTBA and RPA calculations for the first
excited states of the natural parity modes in $^{208}$Pb obtained with
these modified parametrizations.
\end{abstract}


\maketitle

\section{Introduction}
\label{sec:Intr}

Magnetic dipole ($M1$) excitations in the $^{208}$Pb nucleus are the object of
numerous experimental and theoretical investigations for several decades.
From the theoretical point of view, one of the reasons
is the possibility of determining the spin-related parameters
of the residual interaction in the calculations of these excitations within the
random-phase approximation (RPA) or its extended versions.
The calculated energies of the unnatural parity excitations are very sensitive
to the values of the underlying model parameters, in particular, of the parameters
of the Skyrme energy-density functional.
The comparison of these energies with experimental data is the only reliable method
to estimate the spin-related parameters of the model.
Another reason is the fragmentation (spread) of the isovector $M1$ resonance
in $^{208}$Pb which is observed in the experiment (see \cite{Laszewski_1988})
but which is absent in RPA where the isovector $M1$ strength
in this nucleus is concentrated in one state.
The description of this fragmentation requires application
of more complicated models going beyond the RPA framework
(see, e.g., Ref. \cite{Kamerdzhiev_2004} for more details).

Most of the early calculations of the $M1$ excitations in $^{208}$Pb
(see, e.g., Refs.~\cite{Vergados_1971,Ring_1973PLB,Tkachev_1976,Speth_1980,Borzov_1984})
were performed within the RPA, the Tamm-Dancoff approximation or within
the Migdal's Theory of Finite Fermi Systems (TFFS, Ref.~\cite{Migdal_1967})
which in its simplest form used in the applications is equivalent to
the RPA with the zero-range residual interaction.
In the following, the $M1$ modes were investigated within the generalized models
in which the one-particle--one-hole
($1p1h$) RPA configuration space is enlarged by adding
$2p2h$, $1p1h\otimes$phonon or two-phonons configurations
(see, e.g., \cite{Lee_1975,Dehesa_1977,Kamerdzhiev_1984,Cha_1984,Khoa_1986,
Kamerdzhiev_1989,Tselyaev_1989,Kamerdzhiev_1991,Kamerdzhiev_1993a}).
However, the fully self-consistent calculations of the $M1$ excitations in $^{208}$Pb
have been performed so far only within the RPA
(see Refs. \cite{Cao_2009,Vesely_2009,Nesterenko_2010,Cao_2011,Wen_2014,Tselyaev_2019}).

In a broad sense, self-consistency means the use of the same energy-density
functional (EDF) $E[\rho]$ (where $\rho$ is the single-particle density matrix)
for the mean field as well as for the RPA residual interaction.
This decreases the number of the free parameters
of the theory and, in principle, increases its predictive power.
Here we use an EDF of Skyrme type \cite{Bender_2003}.
In a recent paper \cite{Tselyaev_2019}, we have shown that the adequate description
of the low-energy $M1$ excitations in $^{208}$Pb within the self-consistent RPA
based on the Skyrme EDF is possible only if the spin-related parameters of the
known EDF are modified.
By re-tuning these parameters we managed to reproduce within the RPA the experimental
key quantities: energy and the strength of the $1_1^+$ state as well as
the mean energy and the summed strength of the $M1$ resonance in $^{208}$Pb
in the interval 6.6-8.1 MeV. However, as mentioned above, the observed fragmentation
of the isovector $M1$ resonance and its total width in this model are not yet reproduced.

The aim of the present paper is to study the possibility to describe this
fragmentation within the extended self-consistent model including the
$1p1h\otimes$phonon configurations on top of the RPA $1p1h$ configurations.
This extended model is treated within the time blocking approximation (TBA)
which we use actually in its renormalized version (RenTBA, \cite{Tselyaev_2018}).
Full self-consistency is maintained also for the extended treatment.
The method of re-tuning the spin-related parameters
of the Skyrme EDF developed in Ref.~\cite{Tselyaev_2019} is used also for
the RenTBA.

The paper is organized as follows. In Section~\ref{sec:model} the formalism of
RPA and RenTBA is briefly described.
Section~\ref{sec:Det} contains the numerical details and the calculation scheme.
The main results of the paper are presented in Section~\ref{sec:Res}.
In Section~\ref{sec:fine} the fine structure of the $M1$ strength distributions
in $^{208}$Pb and the impact of the single-particle continuum on this structure
are analyzed.
In Section~\ref{sec:frag} the problem of the fragmentation of the isovector $M1$
resonance in $^{208}$Pb is discussed in detail and the necessary condition of
the description of this fragmentation is determined.
In Section~\ref{sec:elex} we present the results of the RenTBA and RPA calculations
of the low-energy electric excitations in $^{208}$Pb
obtained with the use of the modified parametrizations of the Skyrme EDF.
The conclusions are given in the last section.

\section{The model}
\label{sec:model}

Let us start with the RPA eigenvalue equation
\be
  \sum_{34} \Omega^{\mbsu{RPA}}_{12,34}\,Z^{n}_{34}
  =
  \omega^{\vphu}_{n}\,Z^{n}_{12}\,,
\label{rpaeve}
\ee
where $\omega^{\vphu}_{n}$ is the excitation energy, $Z^{n}_{12}$
is the transition amplitude, and the numerical indices ($1,2,3,\ldots$)
stand for the sets of the quantum numbers of some single-particle basis.
In what follows the indices $p$ and $h$ are used to label the states of
the particles and holes in the basis which diagonalizes
the single-particle density matrix $\rho$ and the single-particle Hamiltonian $h$
in the ground state [see Eq.~(\ref{frpa}) below].
The transition amplitudes are normalized by the condition
\be
  \langle\,Z^{n}\,|\,M^{\mbss{RPA}}_{\vphd}|\,Z^{n} \rangle =
  \mbox{sgn}(\omega^{\vphu}_{n})\,,
\label{rpanorm}
\ee
where
\be
M^{\mbsu{RPA}}_{12,34} =
\delta^{\vphu}_{13}\,\rho^{\vphu}_{42} -
\rho^{\vphu}_{13}\,\delta^{\vphu}_{42}
\label{mrpa}
\ee
is the metric matrix in the RPA.

In the self-consistent RPA based on the EDF $E[\rho]$ the RPA matrix
$\Omega^{\mbsu{RPA}}$ is defined by
\be
\Omega^{\mbsu{RPA}}_{12,34} =
h^{\vphu}_{13}\,\delta^{\vphu}_{42} -
\delta^{\vphu}_{13}\,h^{\vphu}_{42} + \sum_{56}
M^{\mbsu{RPA}}_{12,56}\,{V}^{\vphu}_{56,34}\,,
\label{orpa}
\ee
where the single-particle Hamiltonian $h$ and
the amplitude of the residual interaction $V$ are linked
by the relations
\be
h^{\vphu}_{12} = \frac{\delta E[\rho]}{\delta\rho^{\vphu}_{21}}\,,
\qquad
{V}^{\vphu}_{12,34} =
\frac{\delta^2 E[\rho]}
{\delta\rho^{\vphu}_{21}\,\delta\rho^{\vphu}_{34}}\,.
\label{frpa}
\ee

In the TBA, the counterpart of Eq. (\ref{rpaeve}) has the form
\be
  \sum_{34} \Omega^{\mbsu{TBA}}_{12,34}(\omega^{\vphu}_{\nu})\,z^{\nu}_{34}
  = \omega^{\vphu}_{\nu}\,z^{\nu}_{12}\,,
  \label{tbaeve}
\ee
where
\begin{subequations}
\begin{eqnarray}
  \Omega^{\mbsu{TBA}}_{12,34}(\omega)
  &=&
  \Omega^{\mbsu{RPA}}_{12,34}
  +\sum_{56} M^{\mbsu{RPA}}_{12,56}\,\bar{W}^{\vphu}_{56,34}(\omega)
  \,,
\label{omtba}
\\
  \bar{W}^{\vphu}_{12,34}(\omega)
  &=&
  {W}^{\vphu}_{12,34}(\omega) - {W}^{\vphu}_{12,34}(0)
  \,.
\label{wsub}
\end{eqnarray}
\end{subequations}
The matrix $\Omega^{\mbsu{TBA}}(\omega)$ is energy-dependent
due to the matrix $W(\omega)$ which
represents the induced interaction generated by the intermediate
$1p1h\otimes$phonon configurations.
The subtraction of $W(0)$ in Eq.~(\ref{wsub}) serves to avoid changing
the mean-field ground state \cite{Toe88a,Gue93} and to ensure stability
of solutions of the TBA eigenvalue equation (see \cite{Tselyaev_2013}).
The matrix $W(\omega)$ is defined by the equations
\begin{subequations}
\label{eq:Winduc}
\begin{eqnarray}
  {W}^{\vphu}_{12,34}(\omega)
  &=&
  \sum_{c,\;\sigma}\,\frac{\sigma\,{F}^{c(\sigma)}_{12}
  {F}^{c(\sigma)*}_{34}}
  {\omega - \sigma\,\Omega^{\vphu}_{c}}
  \,,
\label{wdef}
\\
  \Omega^{\vphu}_{c}
  &=&
  \ve^{\vphu}_{p'} - \ve^{\vphu}_{h'} + \omega^{\vphu}_{\nu}
  \,,
  \quad \omega^{\vphu}_{\nu}>0
\,,
\label{omcdef}
\end{eqnarray}
\end{subequations}
where $\sigma = \pm 1$, $\,c = \{p',h',\nu\}$ is a combined index for the
$1p1h\otimes$phonon configurations, $\nu$ is the phonon's index,
$\ve^{\vphu}_{p'}$ and $\ve^{\vphu}_{h'}$ are the particle's and hole's
energies, $\omega^{\vphu}_{\nu}$ is the phonon's energy.
The amplitudes ${F}^{c(\sigma)}_{12}$ have only particle-hole matrix elements
${F}^{c(\sigma)}_{ph}$ and ${F}^{c(\sigma)}_{hp}$.
They are defined by the equations
\begin{subequations}
\label{fcdef}
\be
  {F}^{c(-)}_{12}={F}^{c(+)*}_{21},\qquad
  {F}^{c(-)}_{ph}={F}^{c(+)}_{hp}=0\,,
\label{fcrel}
\ee
\be
  {F}^{c(+)}_{ph} =
  \delta^{\vphu}_{pp'}\,g^{\nu}_{h'h}\!-\!
  \delta^{\vphu}_{h'h}\,g^{\nu}_{pp'}
  \,,
\label{fphdef}
\ee
\end{subequations}
where $g^{\nu}_{12}$ is an amplitude of the quasiparticle-phonon interaction.

In the conventional TBA, the phonon's energies $\omega^{\vphu}_{\nu}$
in Eq.~(\ref{omcdef}) and the amplitudes $g^{\nu}_{12}$ in Eq.~(\ref{fphdef})
are determined within the framework of the RPA.
In the non-linear version of the TBA developed in Ref. \cite{Tselyaev_2018},
the phonon's energies $\omega^{\vphu}_{\nu}$ are the solutions of the
TBA equation (\ref{tbaeve}), while the amplitudes $g^{\nu}_{12}$ are
expressed through the transition amplitudes ${z}^{\nu}_{12}$ which are
also the solutions of Eq.~(\ref{tbaeve}), namely
\be
  g^{\nu}_{12} =
  \sum_{34} {V}^{\vphu}_{12,34}\,{z}^{\nu}_{34}
  \,.
\label{gnudef}
\ee

The normalization condition for the transition amplitudes ${z}^{\nu}_{12}$
has the form
\be
(z^{\nu})^2_{\mbsu{RPA}} + (z^{\nu})^2_{\mbsu{CC}} = 1\,,
\label{zmwz2}
\ee
where
\begin{subequations}
\begin{eqnarray}
 (z^{\nu})^2_{\mbsu{RPA}} &=&
  \mbox{sgn}(\omega^{\vphu}_{\nu})\,
  \langle\,{z}^{\nu}\,|\,M^{\mbss{RPA}}_{\vphd}|\,{z}^{\nu} \rangle\,,
\label{zzrpa}\\
 (z^{\nu})^2_{\mbsu{CC}} &=&
  -\mbox{sgn}(\omega^{\vphu}_{\nu})\,
  \langle\,{z}^{\nu}\,|\,W^{\prime}_{\nu}\,|\,{z}^{\nu} \rangle\,,
\label{zzcc}\\
  W^{\prime}_{\nu} &=&
  \biggl(\frac{d\,W(\omega)}{d\,\omega}\biggr)_{\omega\,=\,\omega_{\nu}}.
\label{wnn2}
\end{eqnarray}
\end{subequations}
The terms $(z^{\nu})^2_{\mbsu{RPA}}$ and $(z^{\nu})^2_{\mbsu{CC}}$
represent the contributions of the $1p1h$ components (RPA)
and of the complex configurations (CC) to the norm (\ref{zmwz2}).
The model includes only those TBA phonons that satisfy the condition
%
\begin{equation}
(z^{\nu})^2_{\mbsu{RPA}} > (z^{\nu})^2_{\mbsu{CC}}\,,
\label{czz1}
\end{equation}
which together with Eq. (\ref{zmwz2}) means that
\begin{equation}
(z^{\nu})^2_{\mbsu{RPA}} > \frac{1}{2}\,.
\label{czz2}
\end{equation}
%
The condition (\ref{czz1}) confines the phonon space to the RPA-like phonons
in agreement with the basic model approximations.

The feedback described above renders the phonon space of RenTBA
fully self-consistent.
In the present paper we use the version of this non-linear model
in which the energies $\omega^{\vphu}_{\nu}$ and the amplitudes ${z}^{\nu}_{12}$
entering Eqs. (\ref{omcdef}) and (\ref{gnudef}) (and only in these equations)
are determined from the solutions of the TBA equation (\ref{tbaeve})
in the diagonal approximation. This model is what we call the
renormalized TBA (RenTBA, see \cite{Tselyaev_2018} for more details).

\section{Numerical details and the calculation scheme}
\label{sec:Det}

The equations of RPA and RenTBA were solved within the fully self-consistent
scheme as described in Refs. \cite{Lyutorovich_2015,Lyutorovich_2016,Tselyaev_2016}.
Wave functions and fields were represented on a spherical grid in coordinate space.
The single-particle basis was discretized by imposing the box boundary condition
with the box radius equal to 18~fm. The particle's energies $\ve^{\vphu}_{p}$ were limited
by the maximum value $\ve^{\mbss{max}}_{p} = 100$ MeV.
The non-linear RenTBA equations were solved by means of the iterative procedure.
The phonon space of the first iteration included the RPA phonons with the energies
$\omega^{\vphu}_{n} \leqslant$ 50 MeV and multipolarities $L \leqslant$ 15
of both the electric and magnetic types
which have been selected according to the criterion of collectivity
\be
\langle\,Z^{n}\,|\,V^2|\,Z^{n} \rangle/ \omega^2_n \geqslant 0.05\,,
\label{g2crit}
\ee
see \cite{Tselyaev_2018}.

The field operator $\bfbq$ in the case of the $M1$ excitations was taken in the form
\bea
\bfbq &=& \mu^{\vphu}_N\,\sqrt{\frac{3}{16\pi}}\,
\Bigl\{ (\gamma_n + \gamma_p\,)\,\bfsigma + \bfl
\nonumber\\
&+& \bigl[\,(1-2\xi_s)\,(\gamma_n - \gamma_p\,)\,\bfsigma
- (1-2\xi_{\,l})\,\bfl\,\bigr]\,\tau_3 \Bigr\}
\label{def:qnp}
\eea
where $\bfl$ is the single-particle operator of the angular momentum,
$\bfsigma$ and $\tau_3$ are the spin and isospin Pauli matrices,
respectively (with positive eigenvalue of $\tau_3$ for the neutrons),
$\mu^{\vphu}_N = e\hbar/2m_p c$ is the nuclear magneton, $\gamma_p =
2.793$ and $\gamma_n = -1.913$ are the spin gyromagnetic ratios,
$\xi_s$ and $\xi_{\,l}$ are the renormalization constants.
The nonzero $\xi_s$ and $\xi_{\,l}$ correspond to the effective operator $\bfbq$,
however in the present calculations we used $\xi_{\,l}=0$.
Thus, the reduced probability of the $M1$ excitations $B(M1)$ is defined as
$|\langle\,Z^{n}\,|\,\bfbq \rangle|^2$ in the RPA and as
$|\langle\,z^{\nu}\,|\,\bfbq \rangle|^2$ in the RenTBA.

The Skyrme EDF with the basis parametrizations SKXm \cite{Brown_1998}
and SV-bas \cite{Kluepfel_2009} was used both in RPA and RenTBA.
The nuclear matter parameters for these parametrizations are listed in
Table~\ref{tab:LMP}.
\begin{table}[h!]
\caption{\label{tab:LMP}
Nuclear matter parameters: effective mass $m^*/m$, incompressibility $K_\infty$,
Thomas-Reiche-Kuhn sum rule enhancement factor $\kappa_{\mbss{TRK}}$, and
symmetry energy $a_{\mbss{sym}}$ for two Skyrme-EDF parametrizations:
SKXm \cite{Brown_1998} and SV-bas \cite{Kluepfel_2009}.}
\begin{ruledtabular}
\begin{tabular}{lcccc}
 EDF & $m^*/m$ & $K_{\infty}$ & $\kappa_{\mbss{TRK}}$ & $a_{\mbss{sym}}$ \\
 && (MeV) && (MeV) \\
\hline
 SKXm   & 0.97 & 238 & 0.34 & 31 \\
 SV-bas & 0.90 & 233 & 0.40 & 30 \\
\end{tabular}
\end{ruledtabular}
\end{table}

There are four experimental characteristics of the $M1$ excitations in $^{208}$Pb which
serve as a benchmark in our calculations: energy and excitation probability
of the isoscalar $1^+_1$ state ($E_1=5.84$ MeV with $B_1(M1)=2.0\;\mu^2_N$,
see \cite{Shizuma_2008})
and the mean energy and the summed strength of the isovector $M1$ resonance in
the interval 6.6--8.1 MeV ($E_2=7.4$ MeV with $B_2(M1) = \sum B(M1)$ = 15.3 $\mu^2_N$).
The latter two quantities have been deduced by combining the data from
Refs. \cite{Shizuma_2008,Koehler_1987}.

To reproduce these key characteristics,
the spin-related EDF parameters $W_0$ (spin-orbit strength),
$x_W$ (proton-neutron balance of the spin-orbit term),
$g$ (Landau parameter for isoscalar spin mode),
and $g'$ (Landau parameter for isovector spin mode)
were refitted as explained in \cite{Tselyaev_2019}
while the remaining spin-related parameters of the functional were switched off.
The values of all other parameters of the functional were kept at
the values of the original parametrizations.
The form of the EDF containing all the parameters mentioned above is given
in Ref.~\cite{Tselyaev_2019}.
The spin-orbit parameters $x_W$ and $W_0$ were
refitted to reproduce the experimental value of $B_1(M1)$.
The parameters $g$ and $g'$ enter the terms of the modified Skyrme EDF
which yield the term $V^s$ of the residual interaction $V$ having the form of the
Landau-Migdal ansatz
\be
V^s = C^{\vphu}_{\mbsu{N}} \bigl(\,g\;\bfsigma\cdot\bfsigmap
+ g'\,\bfsigma\cdot\bfsigmap\;\bftau\cdot\bftaup\,\bigr)\,,
\label{vsmigdal}
\ee
where $C^{\vphu}_{\mbsu{N}}$ is the normalization constant.
These parameters allow us to change the calculated energies of the
isoscalar and isovector $1^+$ states.

Note that the parameter $x_W$ was introduced in Refs. \cite{RF95,Sharma_1995}
(with the use of slightly different notations in \cite{RF95})
to regulate the isospin dependence of the spin-orbit potential.
In the most parametrizations of the Skyrme EDF the value $x_W \geqslant 0$ is used
(see, e.g., last two lines of Table~\ref{tab:xwn}).
In particular, the value $x_W = 1$ (frequently used implicitly) corresponds
to the usual Hartree-Fock approximation in the EDF for the two-body spin-orbit
zero-range interaction.
However, in all these cases the value of $B_1(M1)$ in $^{208}$Pb calculated
within the fully self-consistent RPA is much larger than its experimental value
$B_1(M1)_{\mbsu{exp}}$ = 2.0 $\mu^2_N$.
For instance, $B_1(M1)_{\mbsu{RPA}} \approx 10\,B_1(M1)_{\mbsu{exp}}$
for the SLy5 set \cite{Chabanat_1998} even with the use of the effective
$M1$ operator (\ref{def:qnp}).
In Ref.~\cite{Tselyaev_2019} we have shown that one should use the negative
values of $x_W$ to decrease the calculated $B_1(M1)$ up to $B_1(M1)_{\mbsu{exp}}$.
The values of the parameter $W_0$ should be simultaneously increased because
from the set of the refitted parameters $x_W$, $W_0$, $g$, and $g'$,
only the isoscalar combination of the spin-orbit parameters
$C_0^{\nabla J}=-\ffrac{1}{4}(2+x_W)W_0$ have an impact on the ground-state
characteristics of spherical nuclei (see \cite{Tselyaev_2019} for more details).
This combination remains approximately unchanged
in our refitting procedure, so the quality of the description of the ground-state
properties with the use of the original and modified parametrizations of the
Skyrme EDF is approximately the same.

\section{The main results for the {\boldmath $M1$} resonance in $^{208}\mbox{Pb}$}
\label{sec:Res}


The parametrizations obtained in the result of the refitting procedure described above
are SKXm$_{-0.54}$ and SV-bas$_{-0.50}$ for the RPA
and SKXm$_{-0.49}$ and SV-bas$_{-0.44}$ for the RenTBA
(here and in the following the numerical subindex of the modified parametrization
indicates the value of the parameter $x_W$).
The values of the refitted parameters for these sets are shown in Table \ref{tab:xwn}
along with several sets discussed in Sec.~\ref{sec:frag}.
In addition, we used the renormalization constant $\xi_s$ in the field operator
of the $M1$ excitations (\ref{def:qnp}) to fit the isovector $M1$ strength.
The values of this constant for the RPA and RenTBA are also shown in Table \ref{tab:xwn}.
\begin{table}[h!]
\caption{\label{tab:xwn}
Parameters $x_W$, $W_0$, $g$, and $g'$ of the modified Skyrme EDFs
determined on the basis of the parametrizations SKXm \cite{Brown_1998}
and SV-bas \cite{Kluepfel_2009}.
The Landau-Migdal parameters $g$ and $g'$ are normalized to
$C_{\mbsu{N}} = 300$ MeV$\cdot$fm$^3$.
The renormalization constants $\xi_s$ of the field operator of the $M1$ excitations
corresponding to the each parametrization are shown in the last column.
The parameters of the original sets are shown in the last two lines.
}
\begin{ruledtabular}
\begin{tabular}{lccccc}
EDF & $x_W$ & $W_0$ & $g$ & $g'$ & $\xi_s$ \\
&& (MeV$\cdot$fm$^5$) &&&\\
\hline
   SKXm$_{-0.54}$ & $-$0.54 & 226.0 &    $-$0.078 & 0.430 & 0.156 \\
 SV-bas$_{-0.50}$ & $-$0.50 & 213.0 &    $-$0.028 & 0.516 & 0.156 \\
   SKXm$_{-0.49}$ & $-$0.49 & 218.5 & $\hfm$0.108 & 0.930 & 0.085 \\
  SKXm$'_{-0.49}$ & $-$0.49 & 218.5 & $\hfm$0.108 & 0.900 & 0.085 \\
 SKXm$''_{-0.49}$ & $-$0.49 & 218.5 &    $-$0.067 & 0.435 & 0.151 \\
 SV-bas$_{-0.44}$ & $-$0.44 & 204.7 & $\hfm$0.177 & 1.030 & 0.085 \\
SV-bas$'_{-0.44}$ & $-$0.44 & 204.7 & $\hfm$0.177 & 1.460 & 0.085 \\
             SKXm &       0 & 155.9 &           0 &     0 & \\
           SV-bas & $\hfm$0.55 & 124.6 &        0 &     0 & \\
\end{tabular}
\end{ruledtabular}
\end{table}

Note that the set of the phonons in the RenTBA after the renormalization procedure
with the use of the condition (\ref{czz2})
included 123 electric and 83 magnetic phonons for the parametrization SKXm$_{-0.49}$
and 121 electric and 85 magnetic phonons for the parametrization SV-bas$_{-0.44}$.

\begin{figure}[h!]
\begin{center}
\includegraphics*[trim=2cm 6cm 0cm 2cm,clip=true,scale=0.5,angle=0]{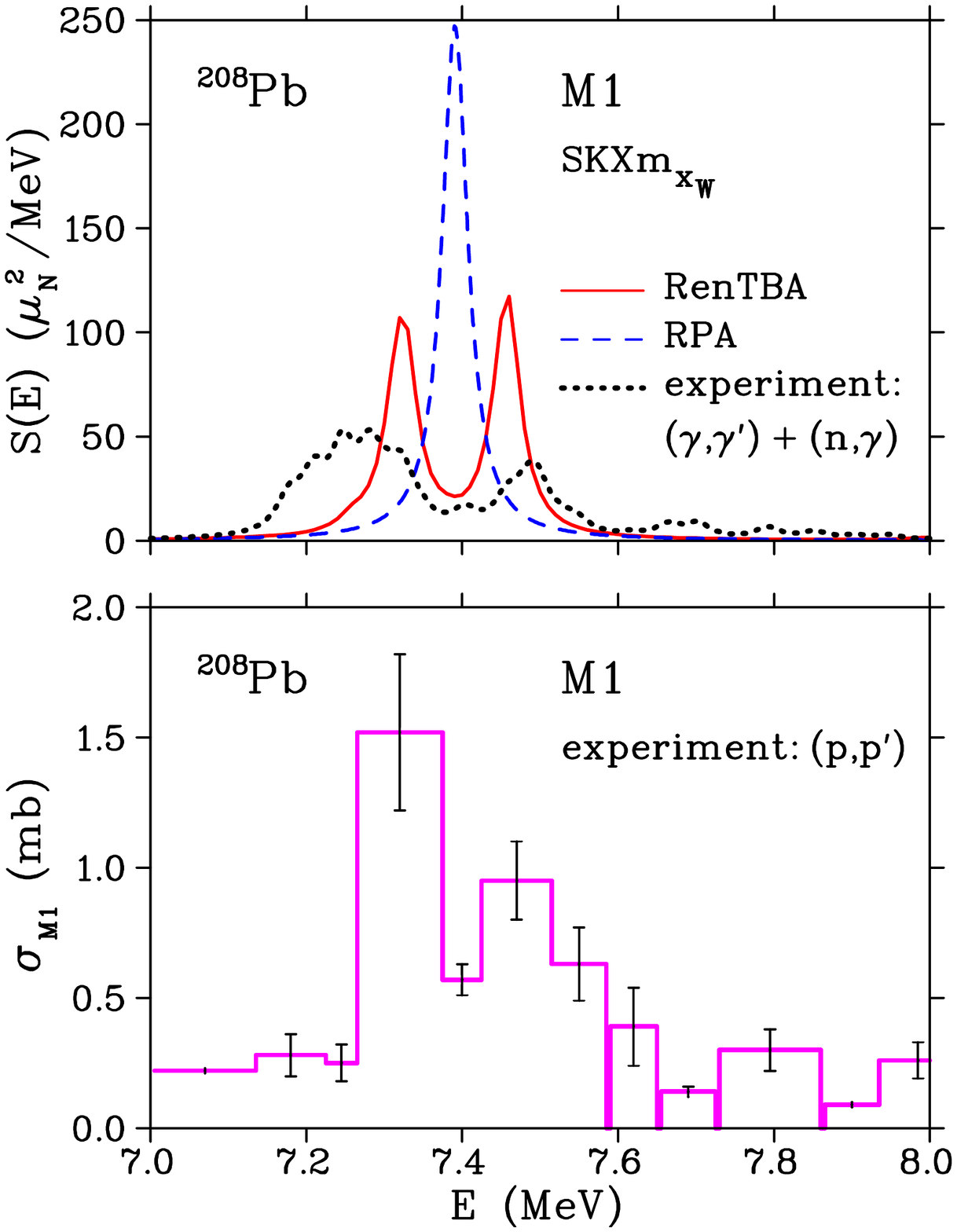}
\end{center}
\caption{\label{fig:m1SKXm}
Upper panel:
strength distributions of the $M1$ excitations in $^{208}$Pb calculated within the
RenTBA with parametrization SKXm$_{-0.49}$ (red solid line)
and within the RPA with parametrization SKXm$_{-0.54}$ (blue dashed line).
The black dotted line represents the strength function (\ref{defse})
obtained from the experimental data \cite{Shizuma_2008,Koehler_1987}.
The smearing parameter $\Delta$ = 20 keV was used. See text for more details.
Lower panel:
the partial $M1$ cross section $\sigma_{M1}$ of the $^{208}$Pb$\,(p,p')$
reaction from Ref.~\cite{Poltoratska12}.
}
\end{figure}

Most of the calculations presented below have been performed within the discrete
versions of RPA and RenTBA that means that the model equations are solved
in the discrete basis representation with the use of the box boundary conditions
for all functions entering these equations.
It is convenient to present these results as well as the experimental data
in the form of the strength functions $S(E)$ obtained by folding the discrete spectra
with a Lorentzian of half-width $\Delta$:
\be
S(E) = \frac{\Delta}{\pi}\sum_{\nu}
\frac{\mbox{sgn}(\omega^{\vphu}_{\nu}) B^{\vphu}_{\nu}(M1)}
{(E - \omega^{\vphu}_{\nu})^2 + \Delta^2}.
\label{defse}
\ee
The results for the modified SKXm parametrizations SKXm$_{-0.49}$ (RenTBA) and
SKXm$_{-0.54}$ (RPA) obtained with $\Delta$ = 20 keV are shown
on the upper panel of Fig.~\ref{fig:m1SKXm}.
The experimental spectra were taken from Refs. \cite{Shizuma_2008}
[$^{208}$Pb$\,(\gamma,\gamma')$ reaction, data below the neutron separation energy
$S(n)=$ 7.37 MeV] and \cite{Koehler_1987}
[$^{207}$Pb$\,(n,\gamma)$ reaction, data above $S(n)$].

The RenTBA, in contrast to the RPA, reproduces the experimental splitting
of the $M1$ resonance into two components separated by the dip near 7.4 MeV.
The quantitative characteristics of this splitting are given in
Table~\ref{tab:inter} in comparison with the experiment.
\begin{table}[h!]
\caption{\label{tab:inter}
The summed strengths $\sum B(M1)$ and the mean energies $\bar{E}$
of the $M1$ excitations calculated within the RenTBA
with parametrization SKXm$_{-0.49}$ in two energy intervals.
The last column contains the Gaussian width $\Gamma$ of the $M1$ strength distribution
calculated in the interval 6.6--8.1 MeV.
The experimental data are taken from Refs. \cite{Shizuma_2008,Koehler_1987}.
}
\begin{ruledtabular}
\begin{tabular}{lcccccc}
interval & \multicolumn{2}{c}{6.60--7.37 MeV} & \multicolumn{2}{c}{7.37--8.10 MeV} && \\
& $\sum B(M1)_{<}$ & $\bar{E}_{<}$ & $\sum B(M1)_{>}$ & $\bar{E}_{>}$ && $\Gamma$ \\
& $(\mu^2_N)$ & (MeV) & $(\mu^2_N)$ & (MeV) && (MeV) \\
\hline
theory     & 7.6 & 7.32 & 7.8 & 7.46 && 0.20 \\
experiment & 9.2 & 7.26 & 6.2 & 7.57 && 0.44 \\
\end{tabular}
\end{ruledtabular}
\end{table}
The experimental summed $M1$ strength in the energy interval below
the neutron threshold $\sum B(M1)_{<}$
is greater than the strength above the threshold $\sum B(M1)_{>}$
by about 50\%, while the respective theoretical values are approximately
equal to each other. Nevertheless, the total theoretical summed $M1$ strength
in the interval 6.6--8.1 MeV is equal to the experimental one according to
the conditions of construction of our modified parametrizations.
The absolute values of the calculated mean energies $\bar{E}_{<}$ and $\bar{E}_{>}$
are close to the experimental values, however the differences
$\Delta\bar{E} = \bar{E}_{>} - \bar{E}_{<}$ are different: the theoretical value
$\Delta\bar{E}_{\mbss{theor}}$ = 0.14 MeV is less than the experimental one
$\Delta\bar{E}_{\mbss{exp}}$ = 0.31 MeV by a factor of two.
To estimate the fragmentation of the $M1$ resonance we have also calculated
the equivalent Gaussian width $\Gamma$ in the interval 6.6--8.1 MeV
both for the experimental and for the theoretical strength distributions.
The results presented in last column of Table~\ref{tab:inter} show that
the total width of the resonance is still underestimated.

Existence of the dip near the neutron separation energy in the experimental
distribution of the $M1$ strength in $^{208}$Pb is generally an uncertain point
because the reliability of the experimental data \cite{Shizuma_2008,Koehler_1987}
goes down in this region. To some extent, the possible existence of this dip
is supported by the more recent data of the $^{208}$Pb$\,(p,p')$ experiment
\cite{Poltoratska12}. The partial $M1$ cross section $\sigma_{M1}$ of this reaction
is shown on the lower panel of Fig. \ref{fig:m1SKXm}.
The dip in energy dependence of $\sigma_{M1}$ near 7.4 MeV
exists though it is less pronounced than for the strength function obtained
from the data \cite{Shizuma_2008,Koehler_1987}.
Note, however, the following. First, the direct comparison of
the $M1$ strength functions $S(E)$ and the cross section $\sigma_{M1}(E)$
is hindered by the fact that they are determined by the different reaction mechanisms.
The distribution of the $B(M1)$ values can be obtained from the cross section
of the $(p,p')$ reaction only within the framework of some model assumptions,
see, e.g., Ref.~\cite{Birkhan_2016}.
Second, the dip near 7.4 MeV is absent in the distribution of $dB(M1)/dE$ deduced in
\cite{Birkhan_2016} from the data of Ref.~\cite{Poltoratska12} and shown
in Fig.~3(b) of Ref.~\cite{Birkhan_2016}. But this fact can be explained by the
different (and quite large) widths of the used energy bins
that corresponds to the large and energy-dependent values of the
smearing parameter $\Delta$ of the strength function (\ref{defse}).

\section{The fine structure of the {\boldmath $M1$} resonance
and the impact of the single-particle continuum}
\label{sec:fine}


To show the fine structure of the theoretical and experimental strength
distributions and to study the role of the single-particle continuum
(which in principle can manifest itself above the neutron separation energy)
we have calculated the $M1$ strength functions in $^{208}$Pb within the continuum RenTBA
with $\Delta$ = 1 keV and 0.1 keV.
The single-particle continuum was included within the response function formalism
according to the method developed in Ref.~\cite{Tselyaev_2016}.
In this approach the strength function $S(E)$ is expressed through
the response function, and the right-hand side of Eq.~(\ref{defse}) is supplemented
with the contribution of the continuum part of the spectrum.
The parametrizations SKXm$_{-0.49}$ and SV-bas$_{-0.44}$ (the latter is discussed
in more detail in Sec.~\ref{sec:frag}) were used.
\begin{figure}[h!]
\begin{center}
\includegraphics*[trim=2cm 6cm 0cm 2cm,clip=true,scale=0.5,angle=0]{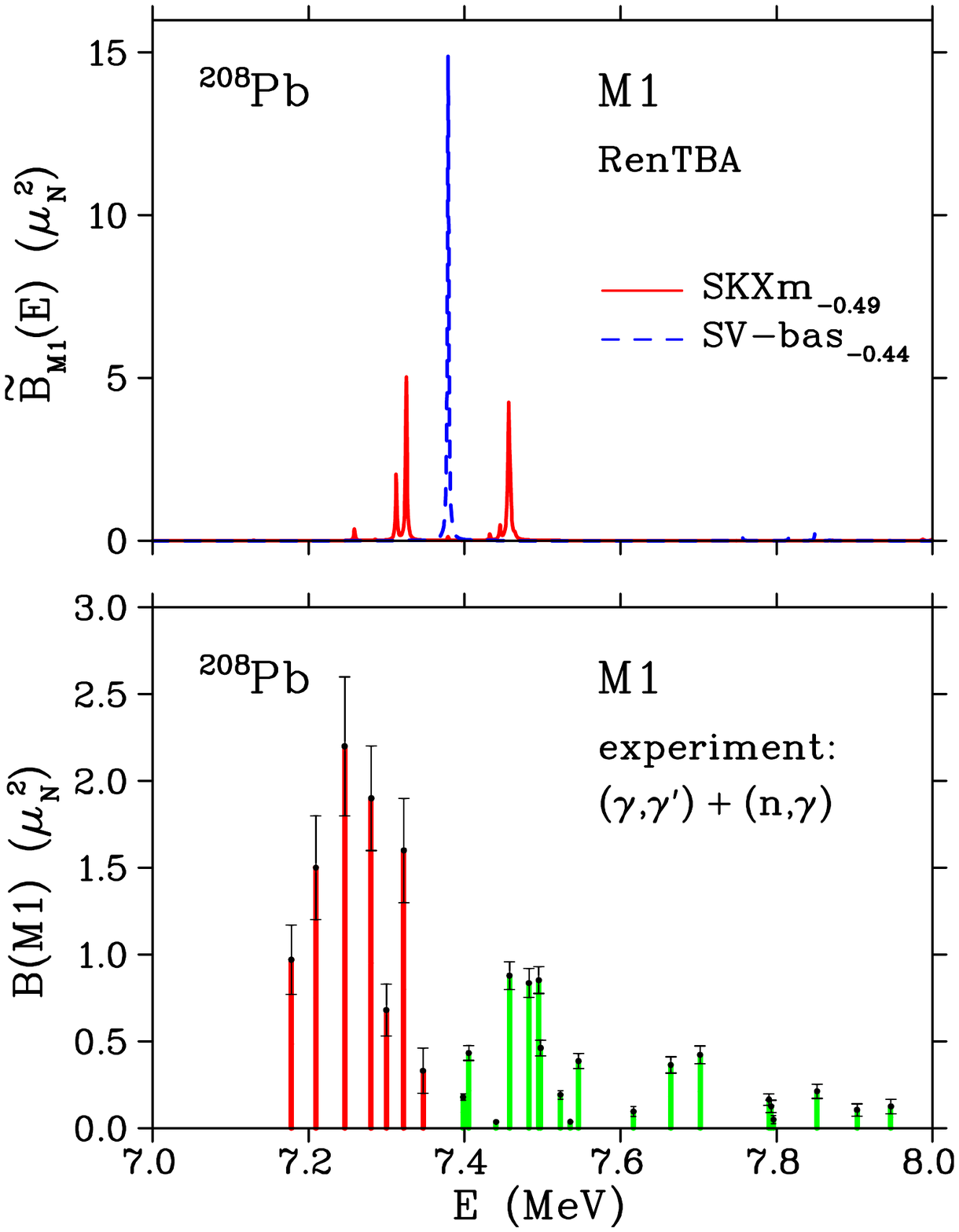}
\end{center}
\caption{\label{fig:m1d1kev}
Upper panel:
strength distributions of the $M1$ excitations in $^{208}$Pb calculated within the
RenTBA with parametrizations SKXm$_{-0.49}$ (red solid line)
and SV-bas$_{-0.44}$ (blue dashed line).
The smearing parameter $\Delta$ = 1 keV was used.
See text for more details.
Lower panel:
experimental distribution of the excitation probabilities $B(M1)$ in $^{208}$Pb
in the interval 7--8 MeV from Refs. \cite{Shizuma_2008}
[$^{208}$Pb$\,(\gamma,\gamma')$ reaction, red vertical lines]
and \cite{Koehler_1987} [$^{207}$Pb$\,(n,\gamma)$ reaction, green vertical lines].
The error bars are indicated by the black lines.
}
\end{figure}
The results for $\Delta$ = 1 keV are shown on the upper panel of
Fig.~\ref{fig:m1d1kev} in terms of the function $\tilde{B}_{M1}(E)$ defined as
\be
\tilde{B}_{M1}(E) = \pi\Delta S(E)\,.
\label{defbt}
\ee
Here we use this function because, as follows from Eq.~(\ref{defse}),
\be
B^{\vphu}_{\nu}(M1) = \lim_{\Delta \rightarrow +0}
\tilde{B}_{M1}(\omega^{\vphu}_{\nu})\,.
\label{blim}
\ee
So, if the $\Delta$ is small, the peak values of the function $\tilde{B}_{M1}(E)$
are close to the excitation probabilities at the peak energies.
Note that Eq.~(\ref{blim}) makes sense only for the states of the discrete spectrum.
However, if the $\Delta$ is greater than the escape width of the quasidiscrete
state in the continuum, the peak value of the function $\tilde{B}_{M1}(E)$
allows us to estimate the integrated strength of the single resonance.

In the RenTBA calculation with the SKXm$_{-0.49}$ set and $\Delta$ = 1 keV,
the fragmentation of two main peaks shown in Fig.~\ref{fig:m1SKXm} for the
strength distributions with $\Delta$ = 20 keV is very small.
This picture does not match the detailed fragmentation structure
of the experimental distribution
composed from data of Refs. \cite{Shizuma_2008,Koehler_1987}
and shown on the lower panel of Fig.~\ref{fig:m1d1kev}.
The $M1$ strength in the interval 7--8 MeV obtained in the RenTBA with
the parametrization SV-bas$_{-0.44}$ is concentrated in one state
without visible fragmentation, as in the case of the RPA.

The lack of fragmentation in the presented RenTBA calculations
can be explained by the limited (though extended as compared to the RPA)
kinds of the correlations included in the model.
There are two natural generalizations of the RenTBA which enable one to
include the additional correlations. First is the model taking into
account the so-called ground-state correlations beyond the RPA.
In Refs. \cite{Kamerdzhiev_1991,Kamerdzhiev_1993a}, it was shown that
the inclusion of the correlations of this type increases the fragmentation
of the $M1$ resonance in $^{208}$Pb.
The second generalization is the replacement of the intermediate
$1p1h\otimes$phonon configurations by two-phonon configurations
according to the scheme suggested in \cite{Tselyaev_2007} and
in analogy with the first versions of the quasiparticle-phonon model
\cite{Soloviev_1992}.
Note that the relative importance of these additional correlations
increases at low energies due to the low level densities as compared
to higher energies.

To analyze the effect of the single-particle continuum we first note
that the theoretical neutron separation energies are equal to
7.30 MeV for the parametrization SKXm$_{-0.49}$ and
7.64 MeV for the parametrization SV-bas$_{-0.44}$.
So, the single peak of the RenTBA strength distribution for the SV-bas$_{-0.44}$ set
shown on the upper panel of Fig.~\ref{fig:m1d1kev} (blue dashed line)
is in the discrete spectrum,
while the main strength of the distribution for the SKXm$_{-0.49}$ set
(red solid line) lies in the continuum.

The effect of the continuum is determined by the values of the escape widths
of the resonances. The full width at half maximum (FWHM) of the single peak
of the strength distribution corresponding to the one or several overlapping
resonances is formed by the escape and spreading widths and by the artificial width
of $2\Delta$ introduced by the smearing parameter. Thus, the FWHM can serve
as an upper bound of the escape width.
The distribution for the parametrization SKXm$_{-0.49}$ shown on the upper panel
of Fig.~\ref{fig:m1d1kev} contains three main peaks with the energies
7.313~MeV, 7.325~MeV, and 7.457~MeV.
These peaks correspond to four states of the discrete RenTBA spectrum with
the energies 7.313~MeV, 7.326~MeV, 7.457~MeV, and 7.459~MeV which exhaust 92\%
of the summed strength of the $M1$ resonance in the interval 6.6--8.1 MeV.
So, we can confine ourselves to analyzing the widths of only these peaks.
The respective values of the FWHM are equal to 2.1 keV for the quasidiscrete states
with $E=$ 7.313~MeV and 7.325~MeV and to 3.4 keV for the resonance with $E=$~7.457 MeV.
The last FWHM value is appreciably greater than $2\Delta$. This is explained by
the fact that the peak with $E=$~7.457 MeV is formed by two overlapping resonances
which correspond to two states of the discrete spectrum mentioned above.

In the calculation with $\Delta$ = 0.1 keV, the widths of the main peaks decrease.
The values of the FWHM for the quasidiscrete states with $E=$ 7.313~MeV and 7.325~MeV
become less than 0.3 keV. The peak with $E=$~7.457 MeV is split into two peaks
separated by the small interval of 2 keV and having the widths which are less
than 1 keV.
Thus the escape widths of the main peaks of the distribution for the SKXm$_{-0.49}$
set are safely less than 1~keV.
These results show that the inclusion of the single-particle continuum has
no appreciable impact in the calculations with $\Delta =$ 20~keV
presented in the paper.

\section{The problem of the fragmentation}
\label{sec:frag}

\begin{figure}[h!]
\begin{center}
\includegraphics*[trim=1cm 0cm 0cm 2cm,clip=true,scale=0.35,angle=90]{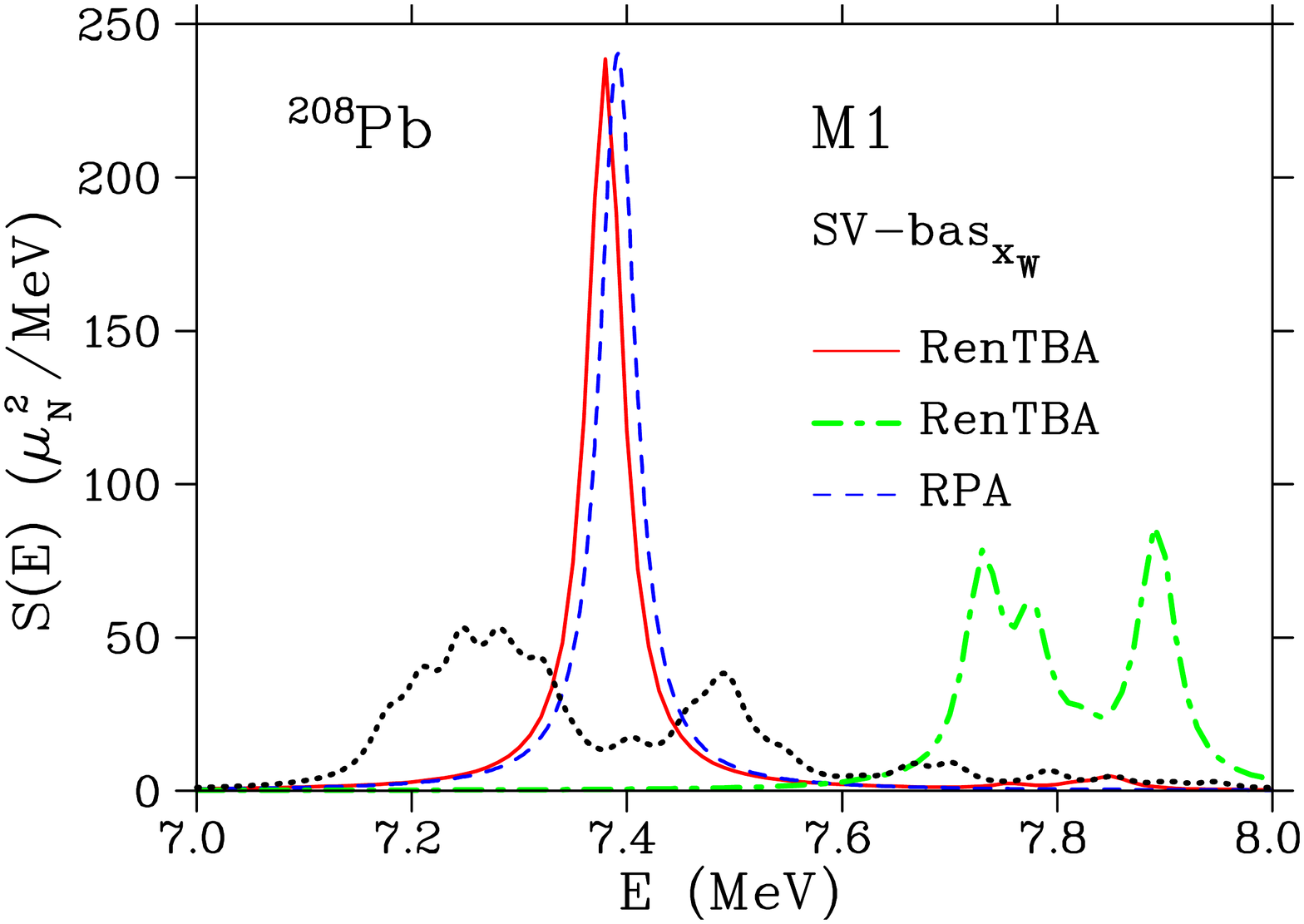}
\end{center}
\caption{\label{fig:m1SVbas}
Strength distributions of the $M1$ excitations in $^{208}$Pb calculated within the
RenTBA with parametrizations SV-bas$_{-0.44}$ (red solid line) and
SV-bas$'_{-0.44}$ (green dashed-dotted line)
and within the RPA with parametrization SV-bas$_{-0.50}$ (blue dashed line)
in comparison with the experiment (black dotted line).
The smearing parameter $\Delta$ = 20 keV was used. See text for more details.
}
\end{figure}

The splitting of the isovector $M1$ resonance in $^{208}$Pb into two main peaks
obtained in RenTBA with the use of the parametrization SKXm$_{-0.49}$
is not a common result for the self-consistent calculations in our approach.
In the typical case, if the EDF parameters $g$ and $g'$ are fitted to reproduce
the experimental energy of the $1_1^+$ state and the mean energy of the $M1$ resonance
in $^{208}$Pb, the fragmentation of the isovector $M1$ resonance is reduced to
the quenching of the main peak without appreciable broadening.
This quenching is compensated by decreasing the renormalization constant $\xi_s$
after which the forms of the RenTBA and RPA $M1$ distributions become close to each other.
This is illustrated in Fig.~\ref{fig:m1SVbas} where we show results for the modified
SV-bas parametrizations.

To clarify the problem, we note that the effects of the fragmentation of the
RPA states in TBA and RenTBA are determined by the energy-dependent
induced interaction ${W}(\omega)$, Eq. (\ref{wdef}).
The fragmentation of the RPA state with the energy $\omega_{\mbss{RPA}}$
is strong if (i) one or more energies $\Omega^{\vphu}_{c}$
of the $1p1h\otimes$phonon configurations in Eqs.~(\ref{eq:Winduc})
are close to the shifted energy $\tilde{\omega}_{\mbss{RPA}}$
(shifted due to the regular contribution of the remaining
$1p1h\otimes$phonon configurations)
and (ii) the respective amplitudes ${F}^{c(+)}_{ph}$ are non-negligible.
In the case of the nucleus $^{208}$Pb, the isovector $M1$ strength in the RPA
is concentrated as a rule in one state with the energy
$\omega_{\mbss{RPA}}(1^+_2)$ (the $1^+_1$ RPA state is isoscalar)
which is formed by two $1p1h$ configurations:
$\pi (1h_{9/2}\otimes 1h^{-1}_{11/2})$ and $\nu (1i_{11/2}\otimes 1i^{-1}_{13/2})$.
So, the $ph$ indices of the amplitudes ${F}^{c(+)}_{ph}$ producing appreciable
fragmentation of the $1^+_2$ RPA state should be one of these two combinations.
Under this condition and according to the selection rules for the $M1$ excitations,
the minimum value of $\Omega^{\vphu}_{c}$ in $^{208}$Pb is determined by the
configuration $c = \{\pi (1h_{9/2}\otimes 3s^{-1}_{1/2})\otimes 5^-_1\}$, that is
\begin{subequations}
\label{eq:omgmin}
\be
\Omega^{\mbsu{min}}_c = \ve^{\pi}_{ph} + \omega(5^-_1)\,,
\label{omgmin}
\ee
where
\be
\ve^{\pi}_{ph} = \ve^{\pi}_p (1h_{9/2}) - \ve^{\pi}_h (3s_{1/2})\,.
\label{ephc}
\ee
\end{subequations}
It turns out that for most Skyrme EDF parametrizations the value
of $\Omega^{\mbsu{min}}_c$ is substantially greater than the mean energy of
the isovector $M1$ resonance in $^{208}$Pb, that is
$\Omega^{\mbsu{min}}_c > 7.4$ MeV. Thus, if the parameters of the EDF
are fitted to reproduce this mean energy, the fragmentation of the isovector
$M1$ resonance is reduced to its quenching as mentioned above.
The parametrization SKXm$_{-0.49}$ is an exception because
the value of $\omega_{\mbss{RenTBA}}(5^-_1)$ comes close the
experimental value which, in turn, yields an
$\Omega^{\mbsu{min}}_c$ close to 7.4 MeV.
This is shown in Table~\ref{tab:omgmin} in comparison with the case
of the SV-bas$_{-0.44}$ parametrization.

\begin{table}
\caption{\label{tab:omgmin}
The values of the particle-hole energies
$\ve^{\pi}_{ph} = \ve^{\pi}_p (1h_{9/2}) - \ve^{\pi}_h (3s_{1/2})$,
the energies of $5^-_1$ phonon, and their sums $\Omega^{\mbsu{min}}_c$,
Eqs. (\ref{eq:omgmin}), in the RenTBA for the parametrizations SKXm$_{-0.49}$
and SV-bas$_{-0.44}$. The experimental values are given in the last line.
}
\begin{ruledtabular}
\begin{tabular}{lccc}
EDF & $\ve^{\pi}_{ph}$ & $\omega(5^-_1)$ & $\Omega^{\mbsu{min}}_c$ \\
& (MeV) & (MeV) & (MeV) \\
\hline
   SKXm$_{-0.49}$ & 4.14 & 3.24 & 7.38 \\
 SV-bas$_{-0.44}$ & 4.27 & 3.55 & 7.82 \\
       experiment & 4.21 & 3.20 & 7.41 \\
\end{tabular}
\end{ruledtabular}
\end{table}

Note that the splitting of the isovector $M1$ resonance shown in Fig. \ref{fig:m1SKXm}
is achieved only in the RenTBA. In conventional TBA, the energies of the phonons
in Eqs.~(\ref{eq:Winduc}) are calculated within the RPA. In the case of
the parametrization SKXm$_{-0.49}$, the energy $\omega_{\mbss{RPA}}(5^-_1)$ = 3.64 MeV
that increases the energy $\Omega^{\mbsu{min}}_c$ and leads to the RPA-like result
in the TBA similar to shown in Fig.~\ref{fig:m1SVbas} by the red solid line.

On the other hand, the fragmentation of the isovector $M1$ resonance in $^{208}$Pb
in itself can be obtained also in the case $\Omega^{\mbsu{min}}_c > 7.4$ MeV
if the isovector $M1$ strength is shifted to higher energies
by increasing the EDF parameter $g'$. This is shown in Fig.~\ref{fig:m1SVbas} for
the parametrization SV-bas$'_{-0.44}$ which is constructed from the set
SV-bas$_{-0.44}$ by changing the parameter $g'$ from 1.03 for SV-bas$_{-0.44}$
to 1.46 for SV-bas$'_{-0.44}$ (however, the set of the phonons in this
illustrative RenTBA calculation for SV-bas$'_{-0.44}$ was used the same
as for SV-bas$_{-0.44}$).
Thus, the simultaneous description of the mean energy of the isovector $M1$ resonance
in $^{208}$Pb and of the fragmentation of this resonance in the self-consistent
calculation is seemingly possible only in rare circumstances as, e.g.,
in case of the parametrization SKXm$_{-0.49}$.

Note that the fragmentation of the isovector $M1$ resonance in $^{208}$Pb was
obtained in the early calculations within the shell model in the $1p1h+2p2h$ space
\cite{Lee_1975} and within the models based on the TFFS \cite{Migdal_1967}
and including the particle-phonon interaction on top of the RPA (see, e.g.,
\cite{Dehesa_1977,Kamerdzhiev_1984,Kamerdzhiev_1989,
Kamerdzhiev_1991,Kamerdzhiev_1993a}).
This result is explained by two reasons.
First, the mean energy of the isovector $M1$ resonance in these calculations was
greater than the experimental value. The shift to higher energies increases the spreading
of the $M1$ strength as was noted in Ref.~\cite{Lee_1975} and is demonstrated
in Fig.~\ref{fig:m1SVbas}.
Second, the phonon energies in the calculations of Refs.
\cite{Kamerdzhiev_1984,Kamerdzhiev_1989,Kamerdzhiev_1991,Kamerdzhiev_1993a}
were fitted to their experimental values that makes the value of $\Omega^{\mbsu{min}}_c$
more close to the mean energy of the isovector $M1$ resonance, see Table \ref{tab:omgmin}.

\begin{figure}
\begin{center}
\includegraphics*[trim=1cm 0cm 0cm 2cm,clip=true,scale=0.35,angle=90]{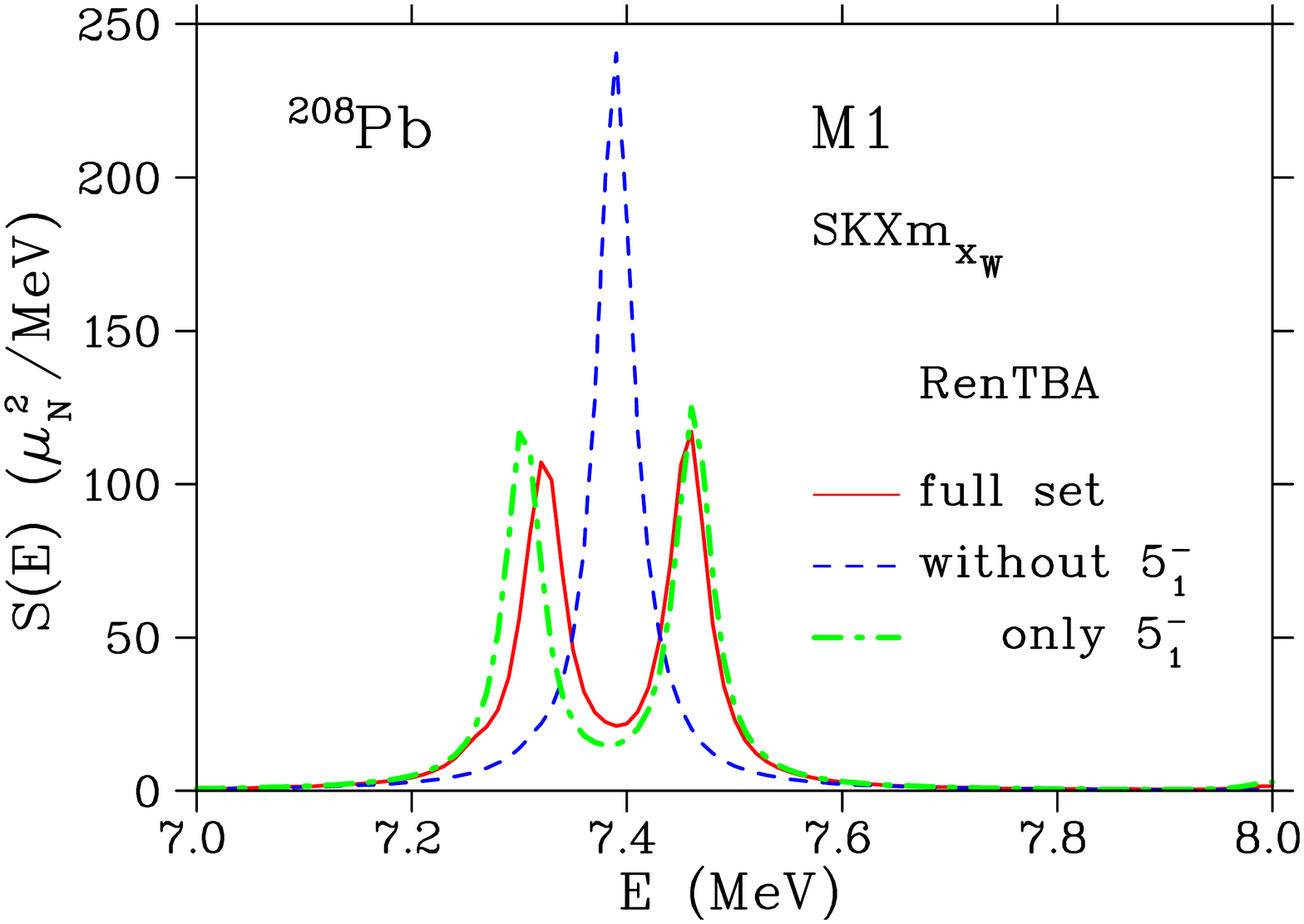}
\end{center}
\caption{\label{fig:m1c5}
Strength distributions of the $M1$ excitations in $^{208}$Pb calculated within the RenTBA
with the full set of the phonons and with parametrization SKXm$_{-0.49}$ (red solid line),
with the set of all the phonons except for the $5^-_1$ phonon
and with parametrization SKXm$'_{-0.49}$ (blue dashed line) and
with the set of the phonons including only the $5^-_1$ phonon
and with parametrization SKXm$''_{-0.49}$ (green dashed-dotted line).
The smearing parameter $\Delta$ = 20 keV was used. See text for more details.
}
\end{figure}

To demonstrate the role of the intermediate $1p1h\otimes$phonon configuration
$\pi (1h_{9/2}\otimes 3s^{-1}_{1/2})\otimes 5^-_1$ in the effect of
the fragmentation under discussion we show in Fig.~\ref{fig:m1c5}
the results of three RenTBA calculations with the use of parametrizations SKXm$_{-0.49}$,
SKXm$'_{-0.49}$, and SKXm$''_{-0.49}$ (see Table \ref{tab:xwn}).
The RenTBA calculation with the use of parametrization SKXm$_{-0.49}$ coincides with
one shown in Fig. \ref{fig:m1SKXm}. In the calculation with the use of SKXm$'_{-0.49}$,
the $5^-_1$ phonon was excluded and the EDF parameter $g'$ was slightly changed to fit
the mean energy of the isovector $M1$ resonance to the experiment.
The calculation with SKXm$''_{-0.49}$ represents the opposite case:
only the $5^-_1$ phonon was included in the phonon basis of the RenTBA and the
EDF parameters $g$ and $g'$ were changed to fit the energy of the $1^+_1$ state
and the mean energy of the isovector $M1$ resonance to the experiment.
The renormalization constant $\xi_s$ was also changed to compensate decreasing
of the quenching of the $M1$ strength. However, the characteristics of the same phonons
(energies, etc.) were the same in all three calculations.
These results show that the splitting of the isovector $M1$ resonance in $^{208}$Pb
is determined in the considered model practically exclusively by the configuration
$\pi (1h_{9/2}\otimes 3s^{-1}_{1/2})\otimes 5^-_1$.
The other $1p1h\otimes$phonon configurations produce only the shift of the $M1$ resonance
and the quenching of the $M1$ strength.

\section{Results for the low-energy electric excitations in $^{208}\mbox{Pb}$}
\label{sec:elex}

In section \ref{sec:frag}, we have shown that the RenTBA using the
modified Skyrme EDF SKXm$_{-0.49}$ gives an energy of the first $5^-$ state in
$^{208}$Pb close to its experimental value. Here we consider the
results of the RenTBA and RPA calculations for the first excited states of
natural parity in $^{208}$Pb with the multipolarity $L$ from 2 to 6 both
for the SKXm$_{-0.49}$ and the SV-bas$_{-0.44}$ parametrizations.
The results are presented in Tables \ref{tab:fst1} and \ref{tab:fst2}.
\begin{table}[h!]
\caption{\label{tab:fst1}
The energies (in MeV) of the first excited states of the natural parity in $^{208}$Pb
calculated within the RenTBA and the RPA with the use of the modified Skyrme EDFs
SKXm$_{-0.49}$ and SV-bas$_{-0.44}$.
The experimental data are taken from Ref.~\cite{Martin_2007}.
}
\begin{ruledtabular}
\begin{tabular}{cccccc}
& \multicolumn{2}{c}{SKXm$_{-0.49}$} & \multicolumn{2}{c}{SV-bas$_{-0.44}$} & \\
$L^{\pi}$ & RenTBA  & RPA & RenTBA  & RPA & experiment \\
\hline
 $2^+_1$ & 4.01 & 4.45 & 4.00 & 4.42 & 4.09 \\
 $3^-_1$ & 2.69 & 2.91 & 2.88 & 3.10 & 2.61 \\
 $4^+_1$ & 4.29 & 4.81 & 4.30 & 4.80 & 4.32 \\
 $5^-_1$ & 3.19 & 3.64 & 3.49 & 3.93 & 3.20 \\
 $6^+_1$ & 4.43 & 5.02 & 4.53 & 5.13 & 4.42 \\
\end{tabular}
\end{ruledtabular}
\end{table}
Note that the RenTBA results have been obtained without use of the diagonal
approximation which is used in the model only for the phonons entering
the intermediate $1p1h\otimes$phonon configurations.
It explains the small difference between the energies of the $5^-_1$ state
listed in Tables \ref{tab:omgmin} (where the diagonal approximation is used)
and \ref{tab:fst1}.

The RenTBA energies calculated with the parametrization SKXm$_{-0.49}$
agree fairly well with the experiment. The deviations for SV-bas$_{-0.44}$
are slightly greater (except for the $4^+_1$ state).
The RPA gives too large energies for both parametrizations.
The energy shift $\omega(\mbox{RPA}) - \omega(\mbox{RenTBA})$
is between 0.2 MeV for the $3^-_1$ state and 0.6 MeV for the $6^+_1$ state.

The situation is the opposite for the excitation probabilities shown in Table~\ref{tab:fst2}.
\begin{table}[h!]
\caption{\label{tab:fst2}
The same as in Table \ref{tab:fst1} but for the excitation probabilities
$B(EL)$ (in units of $e^2$fm$^{2L}$).
}
\begin{ruledtabular}
\begin{tabular}{cccccc}
& \multicolumn{2}{c}{SKXm$_{-0.49}$} & \multicolumn{2}{c}{SV-bas$_{-0.44}$} & \\
$L^{\pi}$ & RenTBA  & RPA & RenTBA  & RPA & experiment \\
\hline
 $2^+_1$ & 2.6$\times$10$^3$ & 3.2$\times$10$^3$ & 2.5$\times$10$^3$ & 3.0$\times$10$^3$ &
 3.2$\times$10$^3$ \\
 $3^-_1$ & 5.6$\times$10$^5$ & 6.4$\times$10$^5$ & 5.8$\times$10$^5$ & 6.4$\times$10$^5$ &
 6.1$\times$10$^5$ \\
 $4^+_1$ & 1.1$\times$10$^7$ & 1.5$\times$10$^7$ & 9.6$\times$10$^6$ & 1.3$\times$10$^7$ &
 1.6$\times$10$^7$ \\
 $5^-_1$ & 1.9$\times$10$^8$ & 2.9$\times$10$^8$ & 2.3$\times$10$^8$ & 3.6$\times$10$^8$ &
 4.5$\times$10$^8$ \\
 $6^+_1$ &
 $\,\,2.6\times$10$^{10}$ & $\,\,3.6\times$10$^{10}$ &
 $\,\,1.4\times$10$^{10}$ & $\,\,2.2\times$10$^{10}$ & $\,\,6.7\times$10$^{10}$ \\
\end{tabular}
\end{ruledtabular}
\end{table}
The RPA results are closer to the experiment as compared to the RenTBA results
(and are in a good agreement with the experiment for $2^+_1$, $3^-_1$,
and $4^+_1$ states). The decrease of the $B(EL)$ values in RenTBA
is caused by the quenching as in the case of the $M1$ excitations.

By construction, the modified parametrizations SKXm$_{-0.49}$ and SV-bas$_{-0.44}$
describe the nuclear ground-state properties within the Skyrme EDF approach
(with approximately the same accuracy as the original parametrizations
SKXm and SV-bas) and reproduce the basic experimental characteristics
of the $M1$ excitations in $^{208}$Pb within the RenTBA.
The results of this section show that the RenTBA with the use of these modified
parametrizations is applicable also to the description of the low-energy
electric excitations in this nucleus.

\section{Conclusions}
\label{sec:conc}

The present paper is a continuation of our recent work \cite{Tselyaev_2019}
in which we investigated the low-energy $M1$ excitations in $^{208}$Pb within
the self-consistent RPA based on the Skyrme energy-density functionals (EDF).
Here we use the extended self-consistent model including the particle-phonon
coupling within the renormalized time blocking approximation (RenTBA, \cite{Tselyaev_2018}).
As in the case of the self-consistent RPA, the description of the basic
experimental characteristics of the $M1$ excitations in $^{208}$Pb
(energy and strength of the $1_1^+$ state as well as
mean energy and summed strength of the isovector $M1$ resonance)
requires refitting some of the spin-related parameters of the Skyrme EDF
within the self-consistent RenTBA.
We have determined several sets of these parameters from this condition.
It has been shown that the observed fragmentation of the isovector $M1$ resonance
in $^{208}$Pb which is absent in all the RPA calculations can be to a certain extent
described within the self-consistent RenTBA. However, this description is not
fully quantitative and is attained only in some cases of the modified
functionals of the Skyrme type.
We have found that the necessary condition to obtain this fragmentation in our model
is the proximity of the energy of the intermediate $1p1h\otimes$phonon
configuration $\pi (1h_{9/2}\otimes 3s^{-1}_{1/2})\otimes 5^-_1$
to the mean energy of the isovector $M1$ resonance in $^{208}$Pb, i.e.
the proximity of the energy of $5^-_1$ phonon to the experimental excitation
energy of the $5^-_1$ state in $^{208}$Pb.
We have also shown that the modified parametrizations of the Skyrme EDF
presented in the paper can be used in the description of the low-energy
electric excitations within the RenTBA.

\begin{acknowledgements}
V.T. is grateful to Prof. V.Yu. Ponomarev for discussions.
Research was carried out using computational resources provided
by Resource Center ``Computer Center of SPbU''.
\end{acknowledgements}

\bibliographystyle{apsrev4-1}
\bibliography{TTT}

\begin{thebibliography}{43}%
\makeatletter
\providecommand \@ifxundefined [1]{%
 \@ifx{#1\undefined}
}%
\providecommand \@ifnum [1]{%
 \ifnum #1\expandafter \@firstoftwo
 \else \expandafter \@secondoftwo
 \fi
}%
\providecommand \@ifx [1]{%
 \ifx #1\expandafter \@firstoftwo
 \else \expandafter \@secondoftwo
 \fi
}%
\providecommand \natexlab [1]{#1}%
\providecommand \enquote  [1]{``#1''}%
\providecommand \bibnamefont  [1]{#1}%
\providecommand \bibfnamefont [1]{#1}%
\providecommand \citenamefont [1]{#1}%
\providecommand \href@noop [0]{\@secondoftwo}%
\providecommand \href [0]{\begingroup \@sanitize@url \@href}%
\providecommand \@href[1]{\@@startlink{#1}\@@href}%
\providecommand \@@href[1]{\endgroup#1\@@endlink}%
\providecommand \@sanitize@url [0]{\catcode `\\12\catcode `\$12\catcode
  `\&12\catcode `\#12\catcode `\^12\catcode `\_12\catcode `\%12\relax}%
\providecommand \@@startlink[1]{}%
\providecommand \@@endlink[0]{}%
\providecommand \url  [0]{\begingroup\@sanitize@url \@url }%
\providecommand \@url [1]{\endgroup\@href {#1}{\urlprefix }}%
\providecommand \urlprefix  [0]{URL }%
\providecommand \Eprint [0]{\href }%
\providecommand \doibase [0]{http://dx.doi.org/}%
\providecommand \selectlanguage [0]{\@gobble}%
\providecommand \bibinfo  [0]{\@secondoftwo}%
\providecommand \bibfield  [0]{\@secondoftwo}%
\providecommand \translation [1]{[#1]}%
\providecommand \BibitemOpen [0]{}%
\providecommand \bibitemStop [0]{}%
\providecommand \bibitemNoStop [0]{.\EOS\space}%
\providecommand \EOS [0]{\spacefactor3000\relax}%
\providecommand \BibitemShut  [1]{\csname bibitem#1\endcsname}%
\let\auto@bib@innerbib\@empty
\bibitem [{\citenamefont {Laszewski}\ \emph {et~al.}(1988)\citenamefont
  {Laszewski}, \citenamefont {Alarcon}, \citenamefont {Dale},\ and\
  \citenamefont {Hoblit}}]{Laszewski_1988}%
  \BibitemOpen
  \bibfield  {author} {\bibinfo {author} {\bibfnamefont {R.~M.}\ \bibnamefont
  {Laszewski}}, \bibinfo {author} {\bibfnamefont {R.}~\bibnamefont {Alarcon}},
  \bibinfo {author} {\bibfnamefont {D.~S.}\ \bibnamefont {Dale}}, \ and\
  \bibinfo {author} {\bibfnamefont {S.~D.}\ \bibnamefont {Hoblit}},\
  }\href@noop {} {\bibfield  {journal} {\bibinfo  {journal} {Phys. Rev. Lett.}\
  }\textbf {\bibinfo {volume} {61}},\ \bibinfo {pages} {1710} (\bibinfo {year}
  {1988})}\BibitemShut {NoStop}%
\bibitem [{\citenamefont {Kamerdzhiev}\ \emph {et~al.}(2004)\citenamefont
  {Kamerdzhiev}, \citenamefont {Speth},\ and\ \citenamefont
  {Tertychny}}]{Kamerdzhiev_2004}%
  \BibitemOpen
  \bibfield  {author} {\bibinfo {author} {\bibfnamefont {S.}~\bibnamefont
  {Kamerdzhiev}}, \bibinfo {author} {\bibfnamefont {J.}~\bibnamefont {Speth}},
  \ and\ \bibinfo {author} {\bibfnamefont {G.}~\bibnamefont {Tertychny}},\
  }\href {\doibase 10.1016/j.physrep.2003.11.001} {\bibfield  {journal}
  {\bibinfo  {journal} {Phys. Rep.}\ }\textbf {\bibinfo {volume} {393}},\
  \bibinfo {pages} {1} (\bibinfo {year} {2004})}\BibitemShut {NoStop}%
\bibitem [{\citenamefont {Vergados}(1971)}]{Vergados_1971}%
  \BibitemOpen
  \bibfield  {author} {\bibinfo {author} {\bibfnamefont {J.~D.}\ \bibnamefont
  {Vergados}},\ }\href@noop {} {\bibfield  {journal} {\bibinfo  {journal}
  {Phys. Lett. B}\ }\textbf {\bibinfo {volume} {36}},\ \bibinfo {pages} {12}
  (\bibinfo {year} {1971})}\BibitemShut {NoStop}%
\bibitem [{\citenamefont {Ring}\ and\ \citenamefont
  {Speth}(1973)}]{Ring_1973PLB}%
  \BibitemOpen
  \bibfield  {author} {\bibinfo {author} {\bibfnamefont {P.}~\bibnamefont
  {Ring}}\ and\ \bibinfo {author} {\bibfnamefont {J.}~\bibnamefont {Speth}},\
  }\href@noop {} {\bibfield  {journal} {\bibinfo  {journal} {Phys. Lett. B}\
  }\textbf {\bibinfo {volume} {44}},\ \bibinfo {pages} {477} (\bibinfo {year}
  {1973})}\BibitemShut {NoStop}%
\bibitem [{\citenamefont {Tkachev}\ \emph {et~al.}(1976)\citenamefont
  {Tkachev}, \citenamefont {Borzov},\ and\ \citenamefont
  {Kamerdzhiev}}]{Tkachev_1976}%
  \BibitemOpen
  \bibfield  {author} {\bibinfo {author} {\bibfnamefont {V.~N.}\ \bibnamefont
  {Tkachev}}, \bibinfo {author} {\bibfnamefont {I.~N.}\ \bibnamefont {Borzov}},
  \ and\ \bibinfo {author} {\bibfnamefont {S.~P.}\ \bibnamefont
  {Kamerdzhiev}},\ }\href@noop {} {\bibfield  {journal} {\bibinfo  {journal}
  {Sov. J. Nucl. Phys.}\ }\textbf {\bibinfo {volume} {24}},\ \bibinfo {pages}
  {373} (\bibinfo {year} {1976})}\BibitemShut {NoStop}%
\bibitem [{\citenamefont {Speth}\ \emph {et~al.}(1980)\citenamefont {Speth},
  \citenamefont {Klemt}, \citenamefont {Wambach},\ and\ \citenamefont
  {Brown}}]{Speth_1980}%
  \BibitemOpen
  \bibfield  {author} {\bibinfo {author} {\bibfnamefont {J.}~\bibnamefont
  {Speth}}, \bibinfo {author} {\bibfnamefont {V.}~\bibnamefont {Klemt}},
  \bibinfo {author} {\bibfnamefont {J.}~\bibnamefont {Wambach}}, \ and\
  \bibinfo {author} {\bibfnamefont {G.~E.}\ \bibnamefont {Brown}},\ }\href@noop
  {} {\bibfield  {journal} {\bibinfo  {journal} {Nucl. Phys. A}\ }\textbf
  {\bibinfo {volume} {343}},\ \bibinfo {pages} {382} (\bibinfo {year}
  {1980})}\BibitemShut {NoStop}%
\bibitem [{\citenamefont {Borzov}\ \emph {et~al.}(1984)\citenamefont {Borzov},
  \citenamefont {Tolokonnikov},\ and\ \citenamefont {Fayans}}]{Borzov_1984}%
  \BibitemOpen
  \bibfield  {author} {\bibinfo {author} {\bibfnamefont {I.~N.}\ \bibnamefont
  {Borzov}}, \bibinfo {author} {\bibfnamefont {S.~V.}\ \bibnamefont
  {Tolokonnikov}}, \ and\ \bibinfo {author} {\bibfnamefont {S.~A.}\
  \bibnamefont {Fayans}},\ }\href@noop {} {\bibfield  {journal} {\bibinfo
  {journal} {Sov. J. Nucl. Phys.}\ }\textbf {\bibinfo {volume} {40}},\ \bibinfo
  {pages} {732} (\bibinfo {year} {1984})}\BibitemShut {NoStop}%
\bibitem [{\citenamefont {Migdal}(1967)}]{Migdal_1967}%
  \BibitemOpen
  \bibfield  {author} {\bibinfo {author} {\bibfnamefont {A.~B.}\ \bibnamefont
  {Migdal}},\ }\href@noop {} {\emph {\bibinfo {title} {Theory of Finite Fermi
  Systems and Application to Atomic Nuclei}}}\ (\bibinfo  {publisher} {Wiley},\
  \bibinfo {address} {New York},\ \bibinfo {year} {1967})\BibitemShut {NoStop}%
\bibitem [{\citenamefont {Lee}\ and\ \citenamefont {Pittel}(1975)}]{Lee_1975}%
  \BibitemOpen
  \bibfield  {author} {\bibinfo {author} {\bibfnamefont {T.-S.~H.}\
  \bibnamefont {Lee}}\ and\ \bibinfo {author} {\bibfnamefont {S.}~\bibnamefont
  {Pittel}},\ }\href@noop {} {\bibfield  {journal} {\bibinfo  {journal} {Phys.
  Rev. C}\ }\textbf {\bibinfo {volume} {11}},\ \bibinfo {pages} {607} (\bibinfo
  {year} {1975})}\BibitemShut {NoStop}%
\bibitem [{\citenamefont {Dehesa}\ \emph {et~al.}(1977)\citenamefont {Dehesa},
  \citenamefont {Speth},\ and\ \citenamefont {Faessler}}]{Dehesa_1977}%
  \BibitemOpen
  \bibfield  {author} {\bibinfo {author} {\bibfnamefont {J.~S.}\ \bibnamefont
  {Dehesa}}, \bibinfo {author} {\bibfnamefont {J.}~\bibnamefont {Speth}}, \
  and\ \bibinfo {author} {\bibfnamefont {A.}~\bibnamefont {Faessler}},\
  }\href@noop {} {\bibfield  {journal} {\bibinfo  {journal} {Phys. Rev. Lett.}\
  }\textbf {\bibinfo {volume} {38}},\ \bibinfo {pages} {208} (\bibinfo {year}
  {1977})}\BibitemShut {NoStop}%
\bibitem [{\citenamefont {Kamerdzhiev}\ and\ \citenamefont
  {Tkachev}(1984)}]{Kamerdzhiev_1984}%
  \BibitemOpen
  \bibfield  {author} {\bibinfo {author} {\bibfnamefont {S.~P.}\ \bibnamefont
  {Kamerdzhiev}}\ and\ \bibinfo {author} {\bibfnamefont {V.~N.}\ \bibnamefont
  {Tkachev}},\ }\href@noop {} {\bibfield  {journal} {\bibinfo  {journal} {Phys.
  Lett. B}\ }\textbf {\bibinfo {volume} {142}},\ \bibinfo {pages} {225}
  (\bibinfo {year} {1984})}\BibitemShut {NoStop}%
\bibitem [{\citenamefont {Cha}\ \emph {et~al.}(1984)\citenamefont {Cha},
  \citenamefont {Schwesinger}, \citenamefont {Wambach},\ and\ \citenamefont
  {Speth}}]{Cha_1984}%
  \BibitemOpen
  \bibfield  {author} {\bibinfo {author} {\bibfnamefont {D.}~\bibnamefont
  {Cha}}, \bibinfo {author} {\bibfnamefont {B.}~\bibnamefont {Schwesinger}},
  \bibinfo {author} {\bibfnamefont {J.}~\bibnamefont {Wambach}}, \ and\
  \bibinfo {author} {\bibfnamefont {J.}~\bibnamefont {Speth}},\ }\href@noop {}
  {\bibfield  {journal} {\bibinfo  {journal} {Nucl. Phys. A}\ }\textbf
  {\bibinfo {volume} {430}},\ \bibinfo {pages} {321} (\bibinfo {year}
  {1984})}\BibitemShut {NoStop}%
\bibitem [{\citenamefont {Khoa}\ \emph {et~al.}(1986)\citenamefont {Khoa},
  \citenamefont {Ponomarev},\ and\ \citenamefont {Vdovin}}]{Khoa_1986}%
  \BibitemOpen
  \bibfield  {author} {\bibinfo {author} {\bibfnamefont {D.~T.}\ \bibnamefont
  {Khoa}}, \bibinfo {author} {\bibfnamefont {V.~Y.}\ \bibnamefont {Ponomarev}},
  \ and\ \bibinfo {author} {\bibfnamefont {A.~I.}\ \bibnamefont {Vdovin}},\
  }\href@noop {} {\bibfield  {journal} {\bibinfo  {journal} {Preprint JINR}\
  }\textbf {\bibinfo {volume} {E4-86-198}} (\bibinfo {year}
  {1986})}\BibitemShut {NoStop}%
\bibitem [{\citenamefont {Kamerdzhiev}\ and\ \citenamefont
  {Tkachev}(1989)}]{Kamerdzhiev_1989}%
  \BibitemOpen
  \bibfield  {author} {\bibinfo {author} {\bibfnamefont {S.~P.}\ \bibnamefont
  {Kamerdzhiev}}\ and\ \bibinfo {author} {\bibfnamefont {V.~N.}\ \bibnamefont
  {Tkachev}},\ }\href@noop {} {\bibfield  {journal} {\bibinfo  {journal} {Z.
  Phys. A}\ }\textbf {\bibinfo {volume} {334}},\ \bibinfo {pages} {19}
  (\bibinfo {year} {1989})}\BibitemShut {NoStop}%
\bibitem [{\citenamefont {Tselyaev}(1989)}]{Tselyaev_1989}%
  \BibitemOpen
  \bibfield  {author} {\bibinfo {author} {\bibfnamefont {V.~I.}\ \bibnamefont
  {Tselyaev}},\ }\href@noop {} {\bibfield  {journal} {\bibinfo  {journal} {Sov.
  J. Nucl. Phys.}\ }\textbf {\bibinfo {volume} {50}},\ \bibinfo {pages} {780}
  (\bibinfo {year} {1989})}\BibitemShut {NoStop}%
\bibitem [{\citenamefont {Kamerdzhiev}\ and\ \citenamefont
  {Tselyaev}(1991)}]{Kamerdzhiev_1991}%
  \BibitemOpen
  \bibfield  {author} {\bibinfo {author} {\bibfnamefont {S.~P.}\ \bibnamefont
  {Kamerdzhiev}}\ and\ \bibinfo {author} {\bibfnamefont {V.~I.}\ \bibnamefont
  {Tselyaev}},\ }\href@noop {} {\bibfield  {journal} {\bibinfo  {journal}
  {Bull. Acad. Sci. USSR, Phys. Ser.}\ }\textbf {\bibinfo {volume} {55}},\
  \bibinfo {pages} {45} (\bibinfo {year} {1991})}\BibitemShut {NoStop}%
\bibitem [{\citenamefont {Kamerdzhiev}\ \emph {et~al.}(1993)\citenamefont
  {Kamerdzhiev}, \citenamefont {Speth}, \citenamefont {Tertychny},\ and\
  \citenamefont {Wambach}}]{Kamerdzhiev_1993a}%
  \BibitemOpen
  \bibfield  {author} {\bibinfo {author} {\bibfnamefont {S.~P.}\ \bibnamefont
  {Kamerdzhiev}}, \bibinfo {author} {\bibfnamefont {J.}~\bibnamefont {Speth}},
  \bibinfo {author} {\bibfnamefont {G.}~\bibnamefont {Tertychny}}, \ and\
  \bibinfo {author} {\bibfnamefont {J.}~\bibnamefont {Wambach}},\ }\href@noop
  {} {\bibfield  {journal} {\bibinfo  {journal} {Z. Phys. A}\ }\textbf
  {\bibinfo {volume} {346}},\ \bibinfo {pages} {253} (\bibinfo {year}
  {1993})}\BibitemShut {NoStop}%
\bibitem [{\citenamefont {Cao}\ \emph {et~al.}(2009)\citenamefont {Cao},
  \citenamefont {Col\`o}, \citenamefont {Sagawa}, \citenamefont {Bortignon},\
  and\ \citenamefont {Sciacchitano}}]{Cao_2009}%
  \BibitemOpen
  \bibfield  {author} {\bibinfo {author} {\bibfnamefont {L.-G.}\ \bibnamefont
  {Cao}}, \bibinfo {author} {\bibfnamefont {G.}~\bibnamefont {Col\`o}},
  \bibinfo {author} {\bibfnamefont {H.}~\bibnamefont {Sagawa}}, \bibinfo
  {author} {\bibfnamefont {P.~F.}\ \bibnamefont {Bortignon}}, \ and\ \bibinfo
  {author} {\bibfnamefont {L.}~\bibnamefont {Sciacchitano}},\ }\href@noop {}
  {\bibfield  {journal} {\bibinfo  {journal} {Phys. Rev. C}\ }\textbf {\bibinfo
  {volume} {80}},\ \bibinfo {pages} {064304} (\bibinfo {year}
  {2009})}\BibitemShut {NoStop}%
\bibitem [{\citenamefont {Vesely}\ \emph {et~al.}(2009)\citenamefont {Vesely},
  \citenamefont {Kvasil}, \citenamefont {Nesterenko}, \citenamefont {Kleinig},
  \citenamefont {Reinhard},\ and\ \citenamefont {Ponomarev}}]{Vesely_2009}%
  \BibitemOpen
  \bibfield  {author} {\bibinfo {author} {\bibfnamefont {P.}~\bibnamefont
  {Vesely}}, \bibinfo {author} {\bibfnamefont {J.}~\bibnamefont {Kvasil}},
  \bibinfo {author} {\bibfnamefont {V.~O.}\ \bibnamefont {Nesterenko}},
  \bibinfo {author} {\bibfnamefont {W.}~\bibnamefont {Kleinig}}, \bibinfo
  {author} {\bibfnamefont {P.-G.}\ \bibnamefont {Reinhard}}, \ and\ \bibinfo
  {author} {\bibfnamefont {V.~Y.}\ \bibnamefont {Ponomarev}},\ }\href@noop {}
  {\bibfield  {journal} {\bibinfo  {journal} {Phys. Rev. C}\ }\textbf {\bibinfo
  {volume} {80}},\ \bibinfo {pages} {031302(R)} (\bibinfo {year}
  {2009})}\BibitemShut {NoStop}%
\bibitem [{\citenamefont {Nesterenko}\ \emph {et~al.}(2010)\citenamefont
  {Nesterenko}, \citenamefont {Kvasil}, \citenamefont {Vesely}, \citenamefont
  {Kleinig}, \citenamefont {Reinhard},\ and\ \citenamefont
  {Ponomarev}}]{Nesterenko_2010}%
  \BibitemOpen
  \bibfield  {author} {\bibinfo {author} {\bibfnamefont {V.~O.}\ \bibnamefont
  {Nesterenko}}, \bibinfo {author} {\bibfnamefont {J.}~\bibnamefont {Kvasil}},
  \bibinfo {author} {\bibfnamefont {P.}~\bibnamefont {Vesely}}, \bibinfo
  {author} {\bibfnamefont {W.}~\bibnamefont {Kleinig}}, \bibinfo {author}
  {\bibfnamefont {P.-G.}\ \bibnamefont {Reinhard}}, \ and\ \bibinfo {author}
  {\bibfnamefont {V.~Y.}\ \bibnamefont {Ponomarev}},\ }\href@noop {} {\bibfield
   {journal} {\bibinfo  {journal} {J. Phys. G: Nucl. Part. Phys.}\ }\textbf
  {\bibinfo {volume} {37}},\ \bibinfo {pages} {064034} (\bibinfo {year}
  {2010})}\BibitemShut {NoStop}%
\bibitem [{\citenamefont {Cao}\ \emph {et~al.}(2011)\citenamefont {Cao},
  \citenamefont {Sagawa},\ and\ \citenamefont {Col\`o}}]{Cao_2011}%
  \BibitemOpen
  \bibfield  {author} {\bibinfo {author} {\bibfnamefont {L.-G.}\ \bibnamefont
  {Cao}}, \bibinfo {author} {\bibfnamefont {H.}~\bibnamefont {Sagawa}}, \ and\
  \bibinfo {author} {\bibfnamefont {G.}~\bibnamefont {Col\`o}},\ }\href@noop {}
  {\bibfield  {journal} {\bibinfo  {journal} {Phys. Rev. C}\ }\textbf {\bibinfo
  {volume} {83}},\ \bibinfo {pages} {034324} (\bibinfo {year}
  {2011})}\BibitemShut {NoStop}%
\bibitem [{\citenamefont {Wen}\ \emph {et~al.}(2014)\citenamefont {Wen},
  \citenamefont {Cao}, \citenamefont {Margueron},\ and\ \citenamefont
  {Sagawa}}]{Wen_2014}%
  \BibitemOpen
  \bibfield  {author} {\bibinfo {author} {\bibfnamefont {P.}~\bibnamefont
  {Wen}}, \bibinfo {author} {\bibfnamefont {L.-G.}\ \bibnamefont {Cao}},
  \bibinfo {author} {\bibfnamefont {J.}~\bibnamefont {Margueron}}, \ and\
  \bibinfo {author} {\bibfnamefont {H.}~\bibnamefont {Sagawa}},\ }\href@noop {}
  {\bibfield  {journal} {\bibinfo  {journal} {Phys. Rev. C}\ }\textbf {\bibinfo
  {volume} {89}},\ \bibinfo {pages} {044311} (\bibinfo {year}
  {2014})}\BibitemShut {NoStop}%
\bibitem [{\citenamefont {Tselyaev}\ \emph {et~al.}(2019)\citenamefont
  {Tselyaev}, \citenamefont {Lyutorovich}, \citenamefont {Speth}, \citenamefont
  {Reinhard},\ and\ \citenamefont {Smirnov}}]{Tselyaev_2019}%
  \BibitemOpen
  \bibfield  {author} {\bibinfo {author} {\bibfnamefont {V.}~\bibnamefont
  {Tselyaev}}, \bibinfo {author} {\bibfnamefont {N.}~\bibnamefont
  {Lyutorovich}}, \bibinfo {author} {\bibfnamefont {J.}~\bibnamefont {Speth}},
  \bibinfo {author} {\bibfnamefont {P.-G.}\ \bibnamefont {Reinhard}}, \ and\
  \bibinfo {author} {\bibfnamefont {D.}~\bibnamefont {Smirnov}},\ }\href
  {https://link.aps.org/doi/1903.01585} {\bibfield  {journal} {\bibinfo
  {journal} {Phys. Rev. C}\ }\textbf {\bibinfo {volume} {99}},\ \bibinfo
  {pages} {064329} (\bibinfo {year} {2019})}\BibitemShut {NoStop}%
\bibitem [{\citenamefont {Bender}\ \emph {et~al.}(2003)\citenamefont {Bender},
  \citenamefont {Heenen},\ and\ \citenamefont {Reinhard}}]{Bender_2003}%
  \BibitemOpen
  \bibfield  {author} {\bibinfo {author} {\bibfnamefont {M.}~\bibnamefont
  {Bender}}, \bibinfo {author} {\bibfnamefont {P.-H.}\ \bibnamefont {Heenen}},
  \ and\ \bibinfo {author} {\bibfnamefont {P.-G.}\ \bibnamefont {Reinhard}},\
  }\href {\doibase 10.1103/Rev.Mod.Phys.75.121} {\bibfield  {journal} {\bibinfo
   {journal} {Rev. Mod. Phys.}\ }\textbf {\bibinfo {volume} {75}},\ \bibinfo
  {pages} {121} (\bibinfo {year} {2003})}\BibitemShut {NoStop}%
\bibitem [{\citenamefont {Tselyaev}\ \emph {et~al.}(2018)\citenamefont
  {Tselyaev}, \citenamefont {Lyutorovich}, \citenamefont {Speth},\ and\
  \citenamefont {Reinhard}}]{Tselyaev_2018}%
  \BibitemOpen
  \bibfield  {author} {\bibinfo {author} {\bibfnamefont {V.}~\bibnamefont
  {Tselyaev}}, \bibinfo {author} {\bibfnamefont {N.}~\bibnamefont
  {Lyutorovich}}, \bibinfo {author} {\bibfnamefont {J.}~\bibnamefont {Speth}},
  \ and\ \bibinfo {author} {\bibfnamefont {P.-G.}\ \bibnamefont {Reinhard}},\
  }\href {\doibase 10.1103/PhysRevC.97.044308} {\bibfield  {journal} {\bibinfo
  {journal} {Phys. Rev. C}\ }\textbf {\bibinfo {volume} {97}},\ \bibinfo
  {pages} {044308} (\bibinfo {year} {2018})}\BibitemShut {NoStop}%
\bibitem [{\citenamefont {Toepffer}\ and\ \citenamefont
  {Reinhard}(1988)}]{Toe88a}%
  \BibitemOpen
  \bibfield  {author} {\bibinfo {author} {\bibfnamefont {C.}~\bibnamefont
  {Toepffer}}\ and\ \bibinfo {author} {\bibfnamefont {P.-G.}\ \bibnamefont
  {Reinhard}},\ }\href@noop {} {\bibfield  {journal} {\bibinfo  {journal} {Ann.
  Phys. (N.Y.)}\ }\textbf {\bibinfo {volume} {181}},\ \bibinfo {pages} {1}
  (\bibinfo {year} {1988})}\BibitemShut {NoStop}%
\bibitem [{\citenamefont {G{\"u}tter}\ \emph {et~al.}(1993)\citenamefont
  {G{\"u}tter}, \citenamefont {Reinhard}, \citenamefont {Wagner},\ and\
  \citenamefont {Toepffer}}]{Gue93}%
  \BibitemOpen
  \bibfield  {author} {\bibinfo {author} {\bibfnamefont {K.}~\bibnamefont
  {G{\"u}tter}}, \bibinfo {author} {\bibfnamefont {P.-G.}\ \bibnamefont
  {Reinhard}}, \bibinfo {author} {\bibfnamefont {K.}~\bibnamefont {Wagner}}, \
  and\ \bibinfo {author} {\bibfnamefont {C.}~\bibnamefont {Toepffer}},\
  }\href@noop {} {\bibfield  {journal} {\bibinfo  {journal} {Ann. Phys.
  (N.Y.)}\ }\textbf {\bibinfo {volume} {225}},\ \bibinfo {pages} {339}
  (\bibinfo {year} {1993})}\BibitemShut {NoStop}%
\bibitem [{\citenamefont {Tselyaev}(2013)}]{Tselyaev_2013}%
  \BibitemOpen
  \bibfield  {author} {\bibinfo {author} {\bibfnamefont {V.~I.}\ \bibnamefont
  {Tselyaev}},\ }\href {\doibase 10.1103/PhysRevC.88.054301} {\bibfield
  {journal} {\bibinfo  {journal} {Phys. Rev. C}\ }\textbf {\bibinfo {volume}
  {88}},\ \bibinfo {pages} {054301} (\bibinfo {year} {2013})}\BibitemShut
  {NoStop}%
\bibitem [{\citenamefont {Lyutorovich}\ \emph {et~al.}(2015)\citenamefont
  {Lyutorovich}, \citenamefont {Tselyaev}, \citenamefont {Speth}, \citenamefont
  {Krewald}, \citenamefont {Gr\"ummer},\ and\ \citenamefont
  {Reinhard}}]{Lyutorovich_2015}%
  \BibitemOpen
  \bibfield  {author} {\bibinfo {author} {\bibfnamefont {N.}~\bibnamefont
  {Lyutorovich}}, \bibinfo {author} {\bibfnamefont {V.}~\bibnamefont
  {Tselyaev}}, \bibinfo {author} {\bibfnamefont {J.}~\bibnamefont {Speth}},
  \bibinfo {author} {\bibfnamefont {S.}~\bibnamefont {Krewald}}, \bibinfo
  {author} {\bibfnamefont {F.}~\bibnamefont {Gr\"ummer}}, \ and\ \bibinfo
  {author} {\bibfnamefont {P.-G.}\ \bibnamefont {Reinhard}},\ }\href@noop {}
  {\bibfield  {journal} {\bibinfo  {journal} {Phys. Lett. B}\ }\textbf
  {\bibinfo {volume} {749}},\ \bibinfo {pages} {292} (\bibinfo {year}
  {2015})}\BibitemShut {NoStop}%
\bibitem [{\citenamefont {Lyutorovich}\ \emph {et~al.}(2016)\citenamefont
  {Lyutorovich}, \citenamefont {Tselyaev}, \citenamefont {Speth}, \citenamefont
  {Krewald},\ and\ \citenamefont {Reinhard}}]{Lyutorovich_2016}%
  \BibitemOpen
  \bibfield  {author} {\bibinfo {author} {\bibfnamefont {N.}~\bibnamefont
  {Lyutorovich}}, \bibinfo {author} {\bibfnamefont {V.}~\bibnamefont
  {Tselyaev}}, \bibinfo {author} {\bibfnamefont {J.}~\bibnamefont {Speth}},
  \bibinfo {author} {\bibfnamefont {S.}~\bibnamefont {Krewald}}, \ and\
  \bibinfo {author} {\bibfnamefont {P.-G.}\ \bibnamefont {Reinhard}},\ }\href
  {ArXiv: http://arxiv.org/abs/1602.00862} {\bibfield  {journal} {\bibinfo
  {journal} {Phys. At. Nucl.}\ }\textbf {\bibinfo {volume} {79}},\ \bibinfo
  {pages} {868} (\bibinfo {year} {2016})}\BibitemShut {NoStop}%
\bibitem [{\citenamefont {Tselyaev}\ \emph {et~al.}(2016)\citenamefont
  {Tselyaev}, \citenamefont {Lyutorovich}, \citenamefont {Speth}, \citenamefont
  {Krewald},\ and\ \citenamefont {Reinhard}}]{Tselyaev_2016}%
  \BibitemOpen
  \bibfield  {author} {\bibinfo {author} {\bibfnamefont {V.}~\bibnamefont
  {Tselyaev}}, \bibinfo {author} {\bibfnamefont {N.}~\bibnamefont
  {Lyutorovich}}, \bibinfo {author} {\bibfnamefont {J.}~\bibnamefont {Speth}},
  \bibinfo {author} {\bibfnamefont {S.}~\bibnamefont {Krewald}}, \ and\
  \bibinfo {author} {\bibfnamefont {P.-G.}\ \bibnamefont {Reinhard}},\ }\href
  {\doibase 10.1103/PhysRevC.94.034306} {\bibfield  {journal} {\bibinfo
  {journal} {Phys. Rev. C}\ }\textbf {\bibinfo {volume} {94}},\ \bibinfo
  {pages} {034306} (\bibinfo {year} {2016})}\BibitemShut {NoStop}%
\bibitem [{\citenamefont {Brown}(1998)}]{Brown_1998}%
  \BibitemOpen
  \bibfield  {author} {\bibinfo {author} {\bibfnamefont {B.~A.}\ \bibnamefont
  {Brown}},\ }\href@noop {} {\bibfield  {journal} {\bibinfo  {journal} {Phys.
  Rev. C}\ }\textbf {\bibinfo {volume} {58}},\ \bibinfo {pages} {220} (\bibinfo
  {year} {1998})}\BibitemShut {NoStop}%
\bibitem [{\citenamefont {Kl{\"{u}}pfel}\ \emph {et~al.}(2009)\citenamefont
  {Kl{\"{u}}pfel}, \citenamefont {Reinhard}, \citenamefont {B{\"{u}}rvenich},\
  and\ \citenamefont {Maruhn}}]{Kluepfel_2009}%
  \BibitemOpen
  \bibfield  {author} {\bibinfo {author} {\bibfnamefont {P.}~\bibnamefont
  {Kl{\"{u}}pfel}}, \bibinfo {author} {\bibfnamefont {P.-G.}\ \bibnamefont
  {Reinhard}}, \bibinfo {author} {\bibfnamefont {T.~J.}\ \bibnamefont
  {B{\"{u}}rvenich}}, \ and\ \bibinfo {author} {\bibfnamefont {J.~A.}\
  \bibnamefont {Maruhn}},\ }\href {\doibase 10.1103/PhysRevC.79.034310}
  {\bibfield  {journal} {\bibinfo  {journal} {Phys. Rev. C}\ }\textbf {\bibinfo
  {volume} {79}},\ \bibinfo {pages} {034310} (\bibinfo {year}
  {2009})}\BibitemShut {NoStop}%
\bibitem [{\citenamefont {Shizuma}\ \emph {et~al.}(2008)\citenamefont
  {Shizuma}, \citenamefont {Hayakawa}, \citenamefont {Ohgaki}, \citenamefont
  {Toyokawa}, \citenamefont {Komatsubara}, \citenamefont {Kikuzawa},
  \citenamefont {Tamii},\ and\ \citenamefont {Nakada}}]{Shizuma_2008}%
  \BibitemOpen
  \bibfield  {author} {\bibinfo {author} {\bibfnamefont {T.}~\bibnamefont
  {Shizuma}}, \bibinfo {author} {\bibfnamefont {T.}~\bibnamefont {Hayakawa}},
  \bibinfo {author} {\bibfnamefont {H.}~\bibnamefont {Ohgaki}}, \bibinfo
  {author} {\bibfnamefont {T.}~\bibnamefont {Toyokawa}}, \bibinfo {author}
  {\bibfnamefont {T.}~\bibnamefont {Komatsubara}}, \bibinfo {author}
  {\bibfnamefont {N.}~\bibnamefont {Kikuzawa}}, \bibinfo {author}
  {\bibfnamefont {A.}~\bibnamefont {Tamii}}, \ and\ \bibinfo {author}
  {\bibfnamefont {H.}~\bibnamefont {Nakada}},\ }\href@noop {} {\bibfield
  {journal} {\bibinfo  {journal} {Phys. Rev. C}\ }\textbf {\bibinfo {volume}
  {78}},\ \bibinfo {pages} {061303(R)} (\bibinfo {year} {2008})}\BibitemShut
  {NoStop}%
\bibitem [{\citenamefont {K{\"{o}}hler}\ \emph {et~al.}(1987)\citenamefont
  {K{\"{o}}hler}, \citenamefont {Wartena}, \citenamefont {Weigmann},
  \citenamefont {Mewissen}, \citenamefont {Poortmans}, \citenamefont
  {Theobald},\ and\ \citenamefont {Raman}}]{Koehler_1987}%
  \BibitemOpen
  \bibfield  {author} {\bibinfo {author} {\bibfnamefont {R.}~\bibnamefont
  {K{\"{o}}hler}}, \bibinfo {author} {\bibfnamefont {J.~A.}\ \bibnamefont
  {Wartena}}, \bibinfo {author} {\bibfnamefont {H.}~\bibnamefont {Weigmann}},
  \bibinfo {author} {\bibfnamefont {L.}~\bibnamefont {Mewissen}}, \bibinfo
  {author} {\bibfnamefont {F.}~\bibnamefont {Poortmans}}, \bibinfo {author}
  {\bibfnamefont {J.~P.}\ \bibnamefont {Theobald}}, \ and\ \bibinfo {author}
  {\bibfnamefont {S.}~\bibnamefont {Raman}},\ }\href@noop {} {\bibfield
  {journal} {\bibinfo  {journal} {Phys. Rev. C}\ }\textbf {\bibinfo {volume}
  {35}},\ \bibinfo {pages} {1646} (\bibinfo {year} {1987})}\BibitemShut
  {NoStop}%
\bibitem [{\citenamefont {Reinhard}\ and\ \citenamefont
  {Flocard}(1995)}]{RF95}%
  \BibitemOpen
  \bibfield  {author} {\bibinfo {author} {\bibfnamefont {P.-G.}\ \bibnamefont
  {Reinhard}}\ and\ \bibinfo {author} {\bibfnamefont {H.}~\bibnamefont
  {Flocard}},\ }\href@noop {} {\bibfield  {journal} {\bibinfo  {journal} {Nucl.
  Phys. A}\ }\textbf {\bibinfo {volume} {584}},\ \bibinfo {pages} {467}
  (\bibinfo {year} {1995})}\BibitemShut {NoStop}%
\bibitem [{\citenamefont {Sharma}\ \emph {et~al.}(1995)\citenamefont {Sharma},
  \citenamefont {Lalazissis}, \citenamefont {K\"onig},\ and\ \citenamefont
  {Ring}}]{Sharma_1995}%
  \BibitemOpen
  \bibfield  {author} {\bibinfo {author} {\bibfnamefont {M.~M.}\ \bibnamefont
  {Sharma}}, \bibinfo {author} {\bibfnamefont {G.}~\bibnamefont {Lalazissis}},
  \bibinfo {author} {\bibfnamefont {J.}~\bibnamefont {K\"onig}}, \ and\
  \bibinfo {author} {\bibfnamefont {P.}~\bibnamefont {Ring}},\ }\href@noop {}
  {\bibfield  {journal} {\bibinfo  {journal} {Phys. Rev. Lett.}\ }\textbf
  {\bibinfo {volume} {74}},\ \bibinfo {pages} {3744} (\bibinfo {year}
  {1995})}\BibitemShut {NoStop}%
\bibitem [{\citenamefont {Chabanat}\ \emph {et~al.}(1998)\citenamefont
  {Chabanat}, \citenamefont {Bonche}, \citenamefont {Haensel}, \citenamefont
  {Meyer},\ and\ \citenamefont {Schaeffer}}]{Chabanat_1998}%
  \BibitemOpen
  \bibfield  {author} {\bibinfo {author} {\bibfnamefont {E.}~\bibnamefont
  {Chabanat}}, \bibinfo {author} {\bibfnamefont {P.}~\bibnamefont {Bonche}},
  \bibinfo {author} {\bibfnamefont {P.}~\bibnamefont {Haensel}}, \bibinfo
  {author} {\bibfnamefont {J.}~\bibnamefont {Meyer}}, \ and\ \bibinfo {author}
  {\bibfnamefont {R.}~\bibnamefont {Schaeffer}},\ }\href@noop {} {\bibfield
  {journal} {\bibinfo  {journal} {Nucl. Phys. A}\ }\textbf {\bibinfo {volume}
  {635}},\ \bibinfo {pages} {231} (\bibinfo {year} {1998})}\BibitemShut
  {NoStop}%
\bibitem [{\citenamefont {Poltoratska}\ \emph {et~al.}(2012)\citenamefont
  {Poltoratska}, \citenamefont {von Neumann-Cosel}, \citenamefont {Tamii},
  \citenamefont {Adachi}, \citenamefont {Bertulani}, \citenamefont {Carter},
  \citenamefont {Dozono}, \citenamefont {Fujita}, \citenamefont {Fujita},
  \citenamefont {Fujita}, \citenamefont {Hatanaka}, \citenamefont {Itoh},
  \citenamefont {Kawabata}, \citenamefont {Kalmykov}, \citenamefont
  {Krumbholz}, \citenamefont {Litvinova}, \citenamefont {Matsubara},
  \citenamefont {Nakanishi}, \citenamefont {Neveling}, \citenamefont {Okamura},
  \citenamefont {Ong}, \citenamefont {\"Ozel-Tashenov}, \citenamefont
  {Ponomarev}, \citenamefont {Richter}, \citenamefont {Rubio}, \citenamefont
  {Sakaguchi}, \citenamefont {Sakemi}, \citenamefont {Sasamoto}, \citenamefont
  {Shimbara}, \citenamefont {Shimizu}, \citenamefont {Smit}, \citenamefont
  {Suzuki}, \citenamefont {Tameshige}, \citenamefont {Wambach}, \citenamefont
  {Yosoi},\ and\ \citenamefont {Zenihiro}}]{Poltoratska12}%
  \BibitemOpen
  \bibfield  {author} {\bibinfo {author} {\bibfnamefont {I.}~\bibnamefont
  {Poltoratska}}, \bibinfo {author} {\bibfnamefont {P.}~\bibnamefont {von
  Neumann-Cosel}}, \bibinfo {author} {\bibfnamefont {A.}~\bibnamefont {Tamii}},
  \bibinfo {author} {\bibfnamefont {T.}~\bibnamefont {Adachi}}, \bibinfo
  {author} {\bibfnamefont {C.~A.}\ \bibnamefont {Bertulani}}, \bibinfo {author}
  {\bibfnamefont {J.}~\bibnamefont {Carter}}, \bibinfo {author} {\bibfnamefont
  {M.}~\bibnamefont {Dozono}}, \bibinfo {author} {\bibfnamefont
  {H.}~\bibnamefont {Fujita}}, \bibinfo {author} {\bibfnamefont
  {K.}~\bibnamefont {Fujita}}, \bibinfo {author} {\bibfnamefont
  {Y.}~\bibnamefont {Fujita}}, \bibinfo {author} {\bibfnamefont
  {K.}~\bibnamefont {Hatanaka}}, \bibinfo {author} {\bibfnamefont
  {M.}~\bibnamefont {Itoh}}, \bibinfo {author} {\bibfnamefont {T.}~\bibnamefont
  {Kawabata}}, \bibinfo {author} {\bibfnamefont {Y.}~\bibnamefont {Kalmykov}},
  \bibinfo {author} {\bibfnamefont {A.~M.}\ \bibnamefont {Krumbholz}}, \bibinfo
  {author} {\bibfnamefont {E.}~\bibnamefont {Litvinova}}, \bibinfo {author}
  {\bibfnamefont {H.}~\bibnamefont {Matsubara}}, \bibinfo {author}
  {\bibfnamefont {K.}~\bibnamefont {Nakanishi}}, \bibinfo {author}
  {\bibfnamefont {R.}~\bibnamefont {Neveling}}, \bibinfo {author}
  {\bibfnamefont {H.}~\bibnamefont {Okamura}}, \bibinfo {author} {\bibfnamefont
  {H.~J.}\ \bibnamefont {Ong}}, \bibinfo {author} {\bibfnamefont
  {B.}~\bibnamefont {\"Ozel-Tashenov}}, \bibinfo {author} {\bibfnamefont
  {V.~Y.}\ \bibnamefont {Ponomarev}}, \bibinfo {author} {\bibfnamefont
  {A.}~\bibnamefont {Richter}}, \bibinfo {author} {\bibfnamefont
  {B.}~\bibnamefont {Rubio}}, \bibinfo {author} {\bibfnamefont
  {H.}~\bibnamefont {Sakaguchi}}, \bibinfo {author} {\bibfnamefont
  {Y.}~\bibnamefont {Sakemi}}, \bibinfo {author} {\bibfnamefont
  {Y.}~\bibnamefont {Sasamoto}}, \bibinfo {author} {\bibfnamefont
  {Y.}~\bibnamefont {Shimbara}}, \bibinfo {author} {\bibfnamefont
  {Y.}~\bibnamefont {Shimizu}}, \bibinfo {author} {\bibfnamefont {F.~D.}\
  \bibnamefont {Smit}}, \bibinfo {author} {\bibfnamefont {T.}~\bibnamefont
  {Suzuki}}, \bibinfo {author} {\bibfnamefont {Y.}~\bibnamefont {Tameshige}},
  \bibinfo {author} {\bibfnamefont {J.}~\bibnamefont {Wambach}}, \bibinfo
  {author} {\bibfnamefont {M.}~\bibnamefont {Yosoi}}, \ and\ \bibinfo {author}
  {\bibfnamefont {J.}~\bibnamefont {Zenihiro}},\ }\href@noop {} {\bibfield
  {journal} {\bibinfo  {journal} {Phys. Rev. C}\ }\textbf {\bibinfo {volume}
  {85}},\ \bibinfo {pages} {041304(R)} (\bibinfo {year} {2012})}\BibitemShut
  {NoStop}%
\bibitem [{\citenamefont {Birkhan}\ \emph {et~al.}(2016)\citenamefont
  {Birkhan}, \citenamefont {Matsubara}, \citenamefont {von Neumann-Cosel},
  \citenamefont {Pietralla}, \citenamefont {Ponomarev}, \citenamefont
  {Richter}, \citenamefont {Tamii},\ and\ \citenamefont
  {Wambach}}]{Birkhan_2016}%
  \BibitemOpen
  \bibfield  {author} {\bibinfo {author} {\bibfnamefont {J.}~\bibnamefont
  {Birkhan}}, \bibinfo {author} {\bibfnamefont {H.}~\bibnamefont {Matsubara}},
  \bibinfo {author} {\bibfnamefont {P.}~\bibnamefont {von Neumann-Cosel}},
  \bibinfo {author} {\bibfnamefont {N.}~\bibnamefont {Pietralla}}, \bibinfo
  {author} {\bibfnamefont {V.~Y.}\ \bibnamefont {Ponomarev}}, \bibinfo {author}
  {\bibfnamefont {A.}~\bibnamefont {Richter}}, \bibinfo {author} {\bibfnamefont
  {A.}~\bibnamefont {Tamii}}, \ and\ \bibinfo {author} {\bibfnamefont
  {J.}~\bibnamefont {Wambach}},\ }\href@noop {} {\bibfield  {journal} {\bibinfo
   {journal} {Phys. Rev. C}\ }\textbf {\bibinfo {volume} {93}},\ \bibinfo
  {pages} {041302(R)} (\bibinfo {year} {2016})}\BibitemShut {NoStop}%
\bibitem [{\citenamefont {Tselyaev}(2007)}]{Tselyaev_2007}%
  \BibitemOpen
  \bibfield  {author} {\bibinfo {author} {\bibfnamefont {V.~I.}\ \bibnamefont
  {Tselyaev}},\ }\href {\doibase 10.1103/Phys.RevC.75.024306} {\bibfield
  {journal} {\bibinfo  {journal} {Phys. Rev. C}\ }\textbf {\bibinfo {volume}
  {75}},\ \bibinfo {pages} {024306} (\bibinfo {year} {2007})}\BibitemShut
  {NoStop}%
\bibitem [{\citenamefont {Soloviev}(1992)}]{Soloviev_1992}%
  \BibitemOpen
  \bibfield  {author} {\bibinfo {author} {\bibfnamefont {V.~G.}\ \bibnamefont
  {Soloviev}},\ }\href@noop {} {\emph {\bibinfo {title} {Theory of Atomic
  Nuclei: Quasiparticles and Phonons}}}\ (\bibinfo  {publisher} {Institute of
  Physics},\ \bibinfo {address} {Bristol and Philadelphia},\ \bibinfo {year}
  {1992})\BibitemShut {NoStop}%
\bibitem [{\citenamefont {Martin}(2007)}]{Martin_2007}%
  \BibitemOpen
  \bibfield  {author} {\bibinfo {author} {\bibfnamefont {M.}~\bibnamefont
  {Martin}},\ }\href {\doibase http://dx.doi.org/10.1016/j.nds.2007.07.001}
  {\bibfield  {journal} {\bibinfo  {journal} {Nuclear Data Sheets}\ }\textbf
  {\bibinfo {volume} {108}},\ \bibinfo {pages} {1583 } (\bibinfo {year}
  {2007})}\BibitemShut {NoStop}%
\end{thebibliography}%
\end{document}